\documentclass[onecolumn,authoryear]{els-mrw} 

\usepackage{amsmath,amssymb,amsfonts,amsthm,makeidx,graphicx}
\usepackage{txfonts}
\usepackage{helvet}
\usepackage{hyperref}

\begin{document}

\chapter{Asteroseismology}
\label{chap: asteroseismology}

\author[1,2]{Dominic M. Bowman}
\author[3]{Lisa Bugnet}

\address[1]{\orgname{School of Mathematics, Statistics and Physics}, \orgdiv{Newcastle University}, \orgaddress{Newcastle upon Tyne, NE1 7RU, UK} \email{dominic.bowman@newcastle.ac.uk}}
\address[2]{\orgname{Institute of Astronomy}, \orgdiv{KU Leuven}, \orgaddress{Celestijnenlaan 200D, 3001 Leuven, Belgium}}
\address[3]{\orgname{Institute of Science and Technology Austria}, \orgdiv{(ISTA)}, \orgaddress{Am Campus 1, Klosterneuburg, Austria}}

\articletag{Chapter Article tagline: update of previous edition,, reprint..}

\maketitle


\begin{abstract}[Abstract] 
Asteroseismology is the study of the interior physics and structure of stars using their pulsations. It is applicable to stars across the Hertzsprung--Russell (HR) diagram and a powerful technique to measure masses, radii and ages, but also directly constrain interior rotation, chemical mixing, and magnetism. This is because a star's self-excited pulsation modes are sensitive to its structure. Asteroseismology generally requires long-duration and high-precision time series data. The method of forward asteroseismic modelling, which is the statistical comparison of observed pulsation mode frequencies to theoretically predicted pulsation frequencies calculated from a grid of models, provides precise constraints for calibrating various transport phenomena. In this introduction to asteroseismology, we provide an overview of its principles, and the typical data sets and methodologies used to constrain stellar interiors. Finally, we present key highlights of asteroseismic results from across the HR~diagram, and conclude with ongoing challenges and future prospects for this ever-expanding field within stellar astrophysics.
\end{abstract}

\begin{keywords}
    Asteroseismology (73); Stellar oscillations (1617); Stellar structures (1631); Stellar evolution (1599); Stellar rotation (1629); Stellar magnetic fields (1610)
\end{keywords}


\begin{BoxTypeA}[box1]{Learning Objectives}
    \begin{enumerate}
    \item The basic principles and fundamentals of asteroseismology.
    \item The common data sets and analysis methods used for analysing stellar pulsations.
    \item The definitions of different types of pulsating stars within the HR~diagram.
    \item The diversity in pulsation frequencies and amplitudes across different types of pulsators.
    \item Key examples of important results from asteroseismology in recent decades.
    \item The remaining major uncertainties and opportunities in stellar structure and evolution theory yet to be fully solved.
    \end{enumerate}
\end{BoxTypeA}

\begin{glossary}[Glossary]

{\bf Eddington limit}: When radiative acceleration acting outwards is balanced by gravity acting inwards, and defines the maximum ratio of $L/M$ a star can have whilst remaining in hydrostatic equilibrium. \\
{\bf Gravity mode}: Type of pulsation mode with a standing wave that has buoyancy as the restoring force. \\
{\bf Hertzsprung--Russell (HR) diagram}: Diagnostic diagram that plots stellar effective temperatures versus their luminosities, and reveals the evolutionary stages for an ensemble of stars. \\
{\bf Hydrostatic equilibrium}: The balance of gravity acting inwards and gas pressure from nuclear fusion acting outwards. \\
{\bf Iterative prewhitening}: Methodology to identify significant pulsation mode frequencies from a time series (e.g. light curve), which utilises Fourier analysis to identify significant pulsation mode frequencies, and then least-squares regression to optimise significant frequencies using a sinusoid model. This model is subtracted from the time series to produce a residual time series and residual frequency spectrum. The process repeats in an iterative fashion until all pulsation mode frequencies have been identified. \\
{\bf Kraft break}: A divider in mass on the main sequence at approximately 1.5~M$_{\odot}$ that separates fast-rotating higher mass stars and slow-rotating low-mass stars, which corresponds to a spectral type of about F5~V.\\
{\bf Main sequence}: The longest part of a star's total lifespan, during which it undergoes hydrogen burning through nuclear fusion in its core. \\
{\bf Mixed mode}: Type of pulsation mode with a standing wave that has the properties of a gravity mode in the deep interior and the properties of a pressure mode in a star's envelope. \\
{\bf Pressure mode}: Type of pulsation mode with a standing (sound; acoustic) wave that has the pressure gradient as the restoring force. \\
{\bf Population~{\sc II} stars}: Old and metal-poor stars often found in the bulges and halos of galaxies that were born in a previous generation of star formation relative to the younger and metal-rich Population~{\sc I} stars born relatively recently in the disks of spiral galaxies. \\
{\bf Red giant star}: An evolved low- or intermediate-mass star that is characterised by being more luminous and redder in colour compared to the Sun, and in the hydrogen-shell or core-helium burning phase of stellar evolution. \\
{\bf Schwarzschild criterion}: Definition for whether a region's dominant energy transport mechanism is convection, with convection occurring if either (i) the radiative temperature gradient is steep enough; or (ii) the adiabatic temperature gradient is shallow enough. \\
{\bf Spherical harmonics}: Solutions for oscillations on the surface of a sphere and includes three quantum numbers: the radial order ($n$) which defines the number of nodes in the radial direction; the angular degree ($\ell$) which defines the number of surface nodal lines; and the azimuthal order ($m$) which defines the number of surface nodal lines that are lines of longitude, such that $|m| \leq \ell$. \\
{\bf Strange mode}: Type of non-linear radial pulsation-like instability caused by the strong
enhancements in opacity within the non-adiabatic partial ionization zones in stars near the Eddington limit (e.g. very massive stars). \\
{\bf Stochastic low-frequency (SLF) variability}: A non-periodic form of variability detected to be seemingly ubiquitous in the light curves of massive stars. \\
{\bf Solar-like oscillator (SLO)}: A group of pulsating stars with stochastically excited pulsations driven by turbulent envelope convection, such that they have the same pulsation excitation mechanism to the Sun. This group predominantly includes solar-type stars and red giant stars. \\
\end{glossary}

\begin{glossary}[Nomenclature and Common Units]
    \begin{tabular}{@{}lp{24pc}@{}}
        M$_{\odot}$     &   Solar mass ($1.99 \times 10^{30}$~kg) \\
        R$_{\odot}$     &   Solar radius ($6.96 \times 10^{8}$~m) \\
        d$^{-1}$        &   cycles-per-day; frequency unit used for stellar pulsation frequencies \\
        $\muup$Hz         &   micro-Hertz; frequency unit used for stellar pulsation frequencies \\
         mmag            &   milli-magnitudes; flux unit used to measure the brightness variations of a star \\
        ppm             &   parts-per-million; flux unit used to measure the brightness variations of a star \\
    \end{tabular}
\end{glossary}


\section{Introduction}
\label{section: intro}

In astrophysics one of the most fundamental parameters of a star is its mass. Stars are generally divided into three main groups based on their birth masses: (i) low-mass stars with masses of $M \lesssim 1.2$~M$_{\odot}$; (ii) intermediate-mass stars with masses of $1.2 \lesssim M \lesssim 8$~M$_{\odot}$; and (iii) massive stars with masses of $M \gtrsim 8$~M$_{\odot}$. These definitions of mass regimes are in part motivated by the fact that a star's hydrostatic equilibrium structure differs significantly among these groups during the longest part of their lifetimes, which is the hydrogen-core burning main-sequence phase. For example, the distinction between low-mass and intermediate-mass stars is motivated by intermediate-mass stars having convective hydrogen-burning cores and predominantly radiative envelopes. Whereas low-mass main-sequence stars, like our Sun, have radiative hydrogen-burning cores and convective envelopes \citep{Kippenhahn_BOOK}. Schematics of the differences in structure of three example stars are shown in Fig.~\ref{figure: structure}. Another important aspect of the three mass-regime definitions is the difference in their evolutionary end products. Massive stars are progenitors of core-collapse supernovae and gamma ray bursts, and leave behind neutron stars and black holes as compact remnants at the ends of their lives \citep{Langer2012}. On the other hand, intermediate- and low-mass stars are progenitors of planetary nebulae and white dwarf stars \citep{Kippenhahn_BOOK}. 

During the longest-lived phase of stellar evolution -- the main sequence -- a star is commonly referred to as a dwarf star since it is relatively small whilst it depletes its supply of hydrogen fuel in its core. In order to maintain hydrostatic equilibrium, the increase in mean molecular weight through nuclear fusion in the hydrogen-burning core results in a gradual increase in radius. However, once the hydrogen is depleted entirely, a star has entered the post-main sequence in which its envelope grows to a much larger size in a relatively short amount of time. Post-main sequence stars are referred to as sub-giants and giants in the low- and intermediate-mass regimes and supergiants in the massive-star regime. Dependent on its mass, there are a number of epochs of nuclear burning phases of chemical elements heavier than hydrogen until it is no longer energetically favourable to continue nuclear fusion. Also, different layers within the star can change between having radiation or convection as the dominant energy transport mechanism following the Schwarzschild criterion \citep{Kippenhahn_BOOK}. This is important since different types of oscillations (i.e. pulsations) are more or less sensitive to radiative and convection zones within a star.

Similar to how the frequency of sound waves generated by a person's vibrating vocal cords depend on the density, temperature and chemical composition of the air they are breathing, the natural resonant frequencies of pulsations in a star depend on its mass, radius, age and chemical composition. Therefore, by measuring the frequencies of a star's pulsations, one is able to constrain such properties but also its interior physical processes, such as which layers are undergoing nuclear burning and their dominant energy transport mechanism. This defines the field of asteroseismology as the study of stellar pulsations for the purpose of improving our understanding of stellar structure and evolution theory within astrophysics. Asteroseismology has rapidly grown to become a mature and advanced field in recent decades, primarily to its success in providing extremely high-precision constraints on stellar properties compared to more traditional methods in astrophysics. \citet{ASTERO_BOOK} provides a complete monograph on asteroseismology aimed at postgraduate students and beyond. Whereas, \citet{Aerts2021a} and \citet{Kurtz2022a} provide extensive and recent reviews of how the theoretical and observational aspects of asteroseismology have progressed in the last two decades, respectively. 

The goal of this chapter on asteroseismology in the Encyclopedia of Astrophysics is to provide a modern entry-point for early-career researchers, such as undergraduate and postgraduate students, into the expanding field of asteroseismology. Here we summarise the basic principles of asteroseismology in Section~\ref{section: pulsations}, its common data sets and methods in Section~\ref{section: methods}, and highlight some key modern results of asteroseismology in the literature from the last two decades in Section~\ref{section: results}. We conclude and provide a summary of the ongoing challenges and future opportunities in Section~\ref{section: conclusions}.


\section{Stellar pulsations}
\label{section: pulsations}

Asteroseismology is a powerful method for constraining the interior physical conditions and transport processes of stars -- see monograph by \citet{ASTERO_BOOK}. At its heart is the principle that the self-excited resonant pulsation modes of a star represent small perturbations to its hydrostatic equilibrium structure. Since stellar pulsation modes directly probe a star's structure, it follows that the fundamental data of asteroseismology are the Eigenfrequencies of a star, which correspond to pulsation modes (i.e. standing waves) each with a unique spherical harmonic geometry and frequency. For non-rotating and non-magnetic stars, the wave functions of a star's pulsations are separable into radial and angular components, with the radial component characterised by a radial order ($n$) defining the number of interior shells acting as nodes, and the angular component characterised by an angular degree ($\ell$), which defines the number of nodal lines on the surface, and an azimuthal order ($m$), which is how many surface nodal lines are lines of longitude such that $|m| \leq \ell$. Thus, it is an important goal of asteroseismology to identify the frequencies of pulsations and their spherical harmonic geometry, and ultimately compare them to theoretical models to constrain a star's physics using a robust statistical comparison methodology --- see review by \citet{Aerts2021a}.

\begin{figure}
\centering
\includegraphics[width=0.32\textwidth]{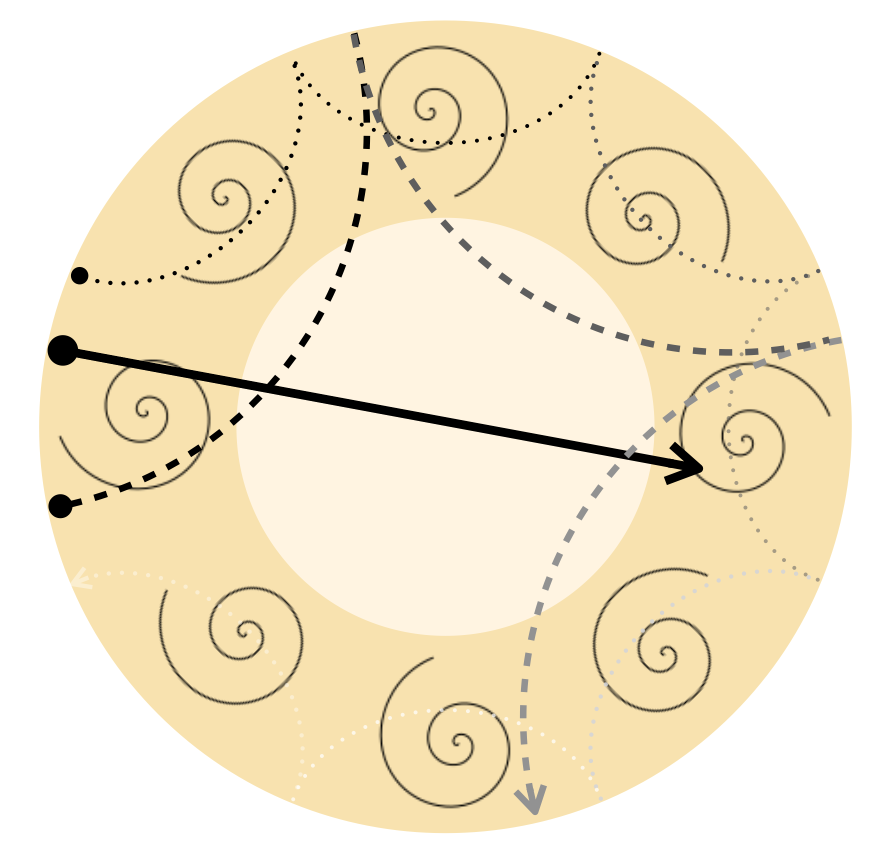}
\includegraphics[width=0.31\textwidth]{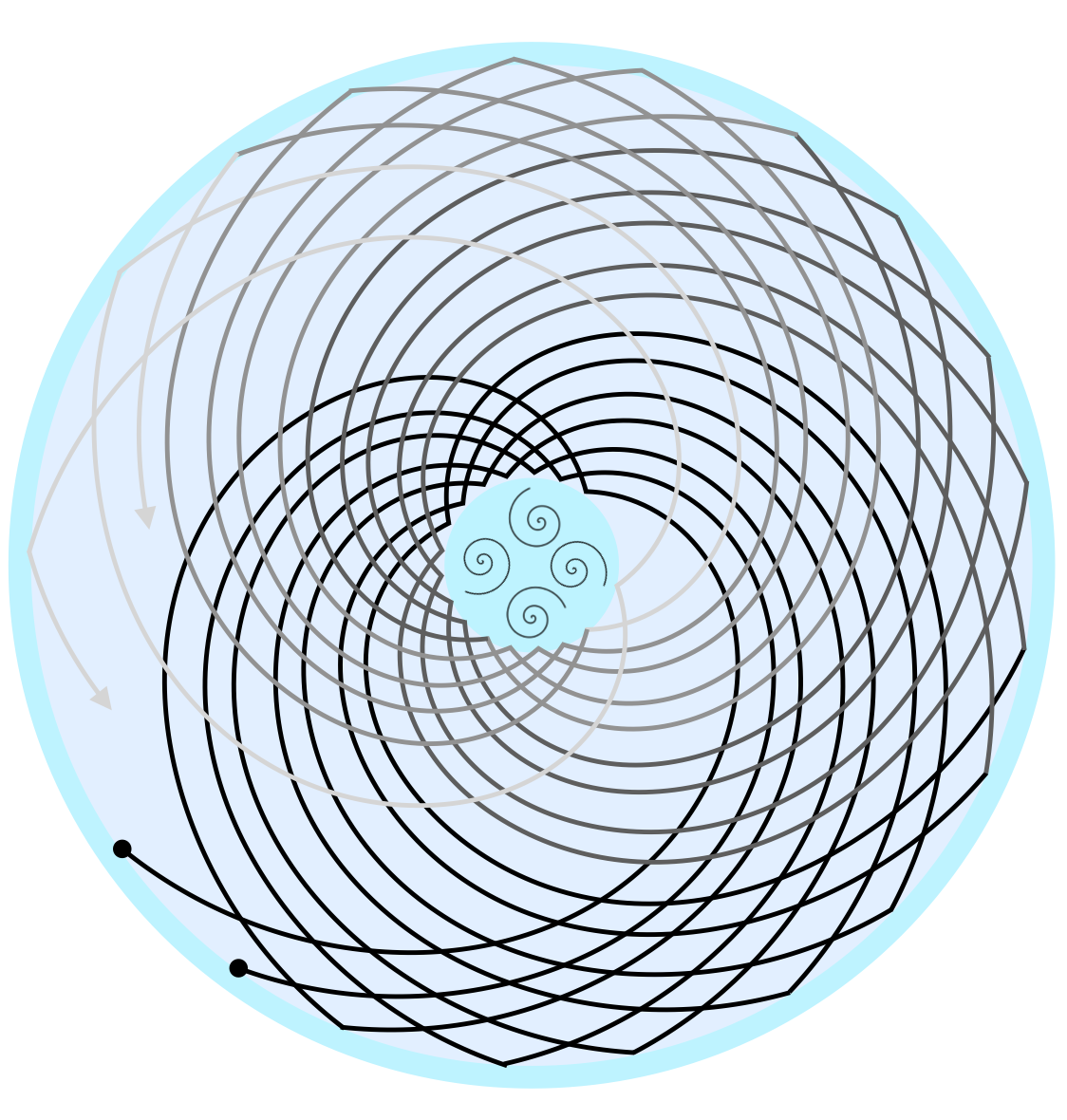}
\includegraphics[width=0.32\textwidth]{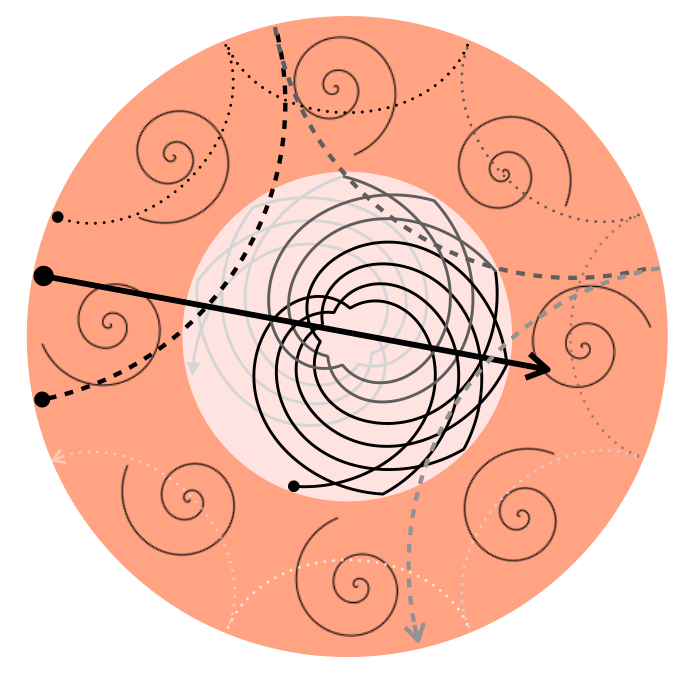}
\caption{Schematics of the different structures of example pulsating stars. Left Panel: Low-mass main-sequence star like the Sun, with the convective envelope indicated by the darker yellow area marked by vortexes, and the radiative interior shown in pale yellow. Each ray-tracing path represents an individual pulsation mode that starts at a black circle. The solid-line arrow indicates a radial (i.e. $\ell=0$) pulsation mode, the dashed-line arrow indicates a quadrupole (i.e. $\ell=2$) mode, and the dotted-line arrow indicates the propagation of a high-radial order high-angular degree mode constrained near the surface. Middle Panel: Schematic of the interior of a more massive star, such as a slowly pulsating B-type (SPB) star, which has a convective core during the main sequence and hosts gravity modes in its radiative envelope. Right Panel: Schematic of the interior of a red giant star hosting mixed pressure-gravity modes that probe its convective envelope and radiative interior.}
\label{figure: structure}
\end{figure}

\begin{figure}
\centering
\includegraphics[width=0.95\textwidth]{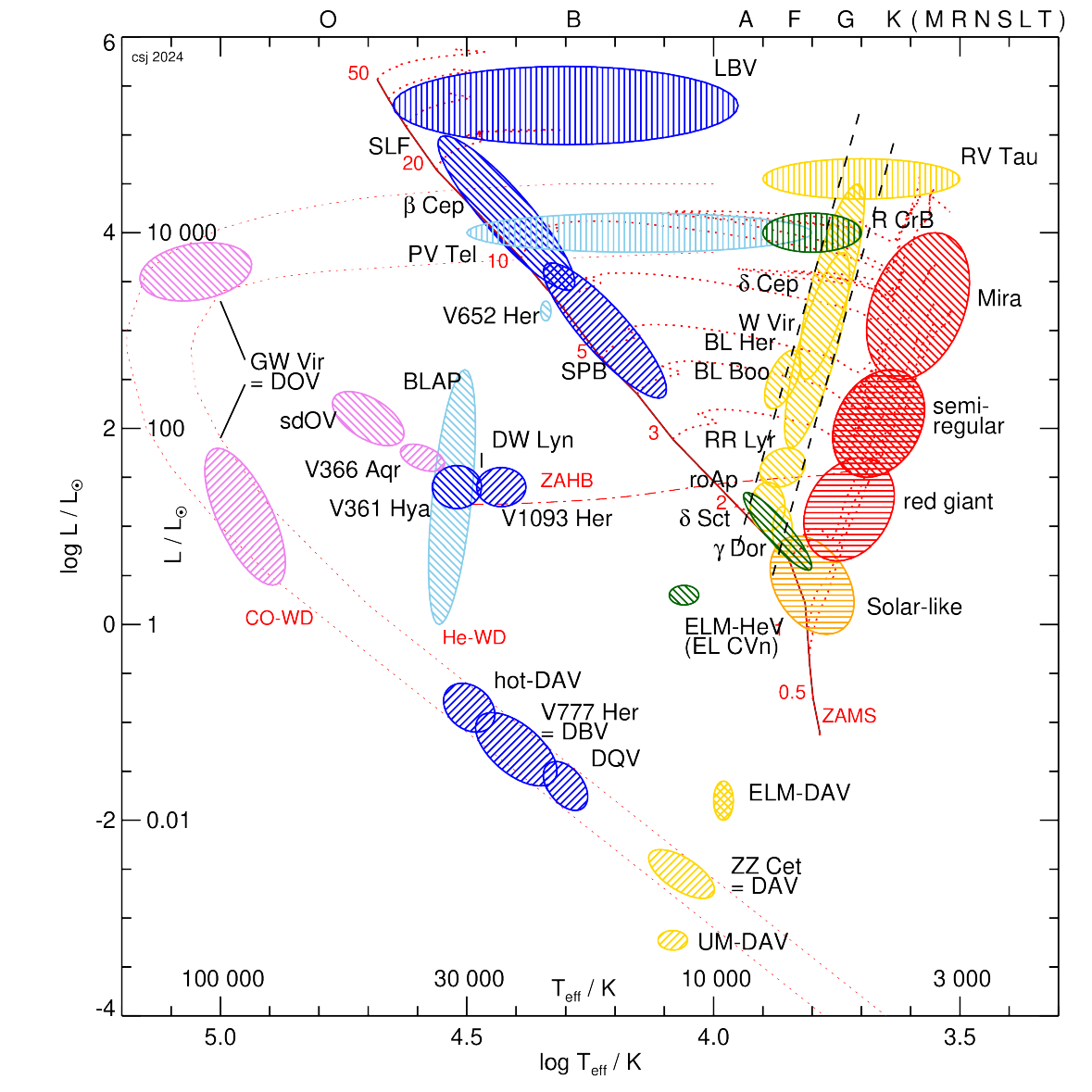}
\caption{Hertzsprung--Russell (HR) diagram of known pulsator classes, courtesy of C. S. Jeffery and based on figure~1 from \citet{Kurtz2022a}, with the original version from \citet{Jeffery2008f}. The hatching of a region denotes the dominant types of pulsation modes for the stars it contains: pressure (p) modes (\textbackslash\textbackslash\textbackslash), gravity (g) modes ($///$), stochastically driven pulsators ($\equiv$), and strange modes ($|||$). Evolutionary tracks of various masses are shown as dotted red lines in units of M$_{\odot}$, with the zero-age main-sequence (ZAMS) line shown as a solid red line. The classical instability strip containing the Cepheid variables at high luminosities and $\delta$~Scuti stars at the intersection with the ZAMS line is delimited by two parallel dashed black lines. The zero-age horizontal branch (ZAHB) for the onset of helium core burning is shown as a dash-dot red line, and the white dwarf cooling track is delimited by two parallel dotted red lines.}
\label{figure: HRD}
\end{figure}

Schematic examples of the pulsation cavities for different types of standing waves for a low-mass main-sequence star like the Sun, a more massive main-sequence star such as a slowly pulsating B-type (SPB) star, and an evolved low-mass star such as a red giant star are shown schematically in Fig.~\ref{figure: structure}, which are set by the physics of their interiors. Consequently, different stellar structures provide the necessary conditions to excite different types of pulsation modes, such that pulsations are expected across the Hertzsprung--Russell (HR) diagram. An asteroseismic HR diagram in which the main types of pulsators are labelled is shown in Fig.~\ref{figure: HRD} (see also \citealt{Jeffery2008f, Kurtz2022a}). The different types of pulsation modes responsible for the plethora of pulsating stars across the HR~diagram are explained in the following subsections.

    \subsection{Pressure modes}
    \label{subsection: pressure}

    Pressure modes are standing sound waves for which the pressure gradient acts as the dominant restoring force, and they can have radial (i.e. $\ell=0$) or non-radial (i.e. $\ell \geq 1$) spherical harmonic geometries. For radial pressure modes, since there is no angular dependence, it is the entire radius of a star that acts as the pulsation cavity. Whereas non-radial pressure modes can only penetrate to a certain depth beneath the stellar surface, which is determined by the local adiabatic sound speed, $c_{\rm s}$, defined as:
    \begin{align}
        \label{equation: sound speed}
        c_{\rm s} = \sqrt{\gamma \frac{P}{\rho}} ~ , 
    \end{align}
    \noindent where $\gamma$ is the adiabatic index, $P$ is the pressure, and $\rho$ is the density of the gas. The pulsation cavity of a pressure mode is related to the requirement of it having a frequency higher than the Lamb frequency, $S_{\ell}$, which is defined as:
    \begin{align}
        \label{equation: Lamb frequency}
        S_{\ell} = \frac{\ell(\ell+1) \, c_{\rm s}^2}{r^2} ~ .
    \end{align}
    \noindent Rearranging Eqn.~(\ref{equation: Lamb frequency}) for radius, $r$, defines the so-called `turning radius' of a non-radial pressure mode. This is the deepest radial coordinate a non-radial pressure mode reaches if excited at the stellar surface. Thus, pressure modes with higher values of $\ell$ have shallower pulsation cavities since the turning radius is proportional to $\sqrt{\ell(\ell+1)}$, which means that higher-$\ell$ pressure modes are more sensitive to the near-surface layers within a star as shown in the schematic in the left panel of Fig.~\ref{figure: structure}.

    For pressure modes with relatively high-radial orders such that $n >> \ell$, with typical values being $n \in \{5,...,15\}$ and $\ell \in \{0,1,2\}$, the frequencies follow an asymptotic approximation such that pressure modes of consecutive radial order ($n$) with the same angular degree ($\ell$) and azimuthal order ($m$) are equally spaced in frequency. The frequency difference within this regular `comb' of frequencies defines the large frequency separation:
    \begin{align}
        \label{equation: delta nu} 
        \Delta\nu = \left( 2 \int_{0}^{R} \frac{{\rm d}r}{c_{\rm s}} \right)^{-1} ~ ,
    \end{align}
    \noindent which is a measure of the mean density of a star since it is related to the sound crossing time of the star.

    Another useful observable to determine when analysing stars with high-radial order pressure modes is the frequency of maximum power, which is often denoted as $\nu_{\rm max}$. Although there is some uncertainty as to how specific radial-order pressure modes are excited by turbulent convection in the envelope of a star like the Sun, a roughly Gaussian envelope of energy is typically injected into a specific frequency regime for solar-like oscillators (SLOs), which includes solar-type main-sequence stars, sub-giant and red giant stars (see reviews by \citealt{Chaplin2013c, Hekker2017a, Garcia_R_2019}). The power excess thus can be approximated by a Gaussian with a centroid corresponding to $\nu_{\rm max}$. As stars evolve, such as when a star ascends the red giant branch in the HR~diagram whilst hydrogen-shell burning, the value of $\nu_{\rm max}$ moves to lower frequencies, driven by a star's increasing radius. It is expected from theory and seen in observations that $\nu_{\rm max}$ follows the relationship: $\nu_{\rm max} \propto \nu_{\rm ac} \propto gT_{\rm eff}^{-1/2}$, where $g$ is the surface gravity, $T_{\rm eff}$ is the effective temperature of the star, and $\nu_{\rm ac}$ is the acoustic cut-off frequency, which is the highest frequency pressure mode that can be supported within a star given its mass and radius. Given this proportionality relationship and by measuring the Sun's $\nu_{\rm max}$ value from the frequency spectrum of its light curve, one can derive:
    \begin{align}
        \nu_{\rm max} = \frac{M/{\rm M}_{\odot}}{(R/{\rm R}_{\odot})^2 \sqrt{T_{\rm eff}/T_{\rm eff,\odot}}} \, \nu_{\rm max, \odot} ~ ,
        \label{equation: nu-max}
    \end{align}
    \noindent where $\nu_{\rm max, \odot} = 3050$~$\muup$Hz and $T_{\rm eff, \odot} = 5772$~K are the solar values. However, this assumes that a star has the same structure as the Sun \citep{Hekker2020c}.
    
\begin{figure}
    \centering
    \includegraphics[width=0.49\textwidth]{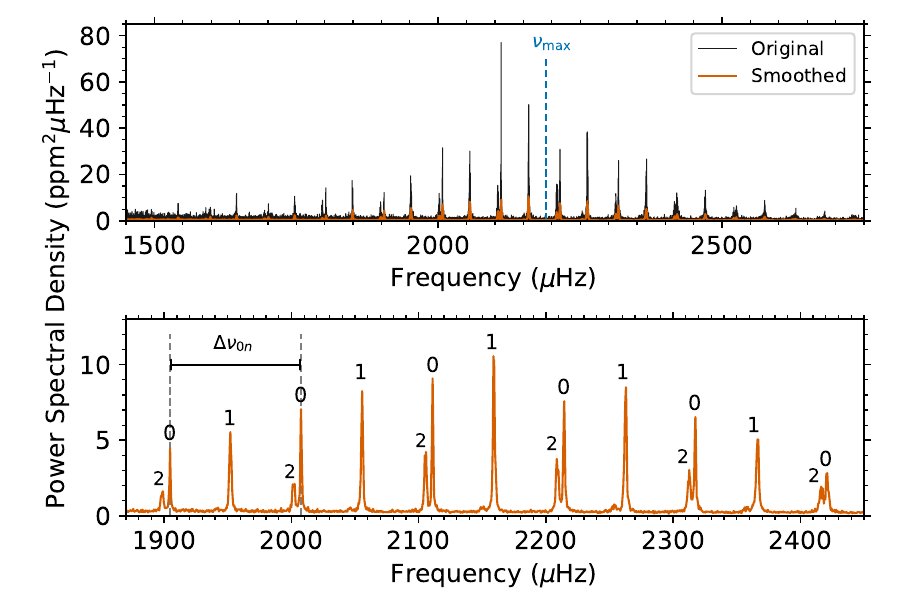}
    \includegraphics[width=0.49\textwidth]{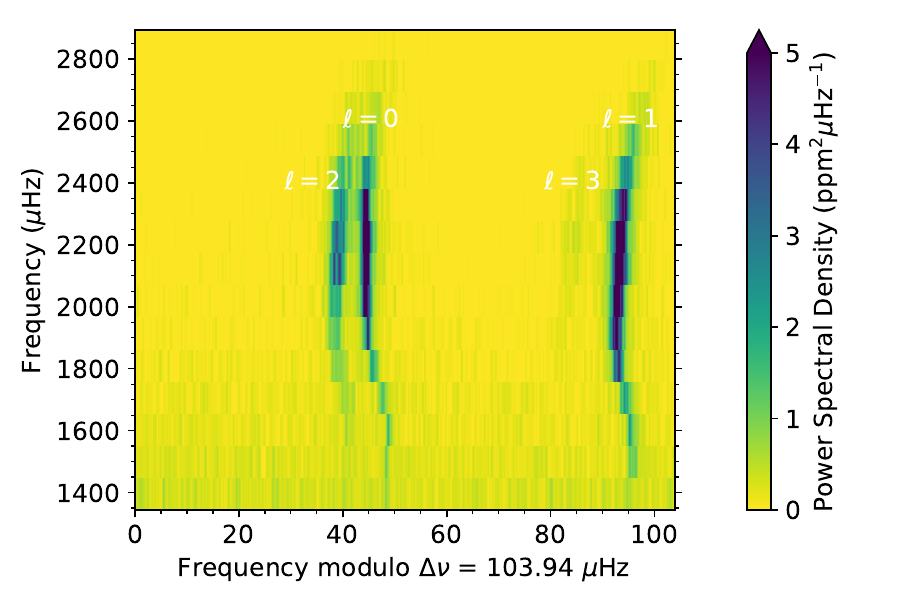}
    \caption{Left Panels: Power spectral density (PSD) spectrum of the solar ``twin'' star 16~Cyg~A observed by the {\it Kepler} mission, which is defined as the squared-amplitude of the Fourier transform per unit frequency of the light curve (an example light curve of a solar-like oscillator is shown in Fig.~\ref{figure: light curves}). The top-left panel shows the frequency regime containing all the significant pressure modes. The bottom-left panel shows a zoom-in of the smoothed PSD spectrum with the angular degree, $\ell$, of individual pulsation modes labelled, as well as the large frequency separation, $\Delta\nu = 103.94$~$\muup$Hz, and frequency of maximum power, $\nu_{\rm max}$, at a frequency of about 2200~$\muup$Hz. Right Panel: {\'e}chelle diagram of 16~Cyg~A with the dominant $\ell = {0,1,2}$ and barely visible $\ell=3$ ridges of pulsation modes separated by $\Delta\nu$.}
    \label{figure: 16 Cyg}
    \end{figure}

    The regular pattern of high-radial order pressure modes equally spaced in frequency and centered on a particular frequency results in measurable values of $\Delta\nu$ and $\nu_{\rm max}$, respectively. These are most readily apparent in the frequency spectra of SLOs including the Sun and red giant stars. An example of the frequency spectrum of the SLO star 16~Cyg~A is shown in Fig.~\ref{figure: 16 Cyg}, in which the individual pulsation modes are labelled, as well as the observables $\Delta\nu$ and $\nu_{\rm max}$. The relationship in Eqn.~(\ref{equation: nu-max}) demonstrates that younger stars have higher values of $\nu_{\rm max}$ to older stars of the same mass. This means that 16~Cyg~A is an older star compared to the Sun given it has a comparable mass of $1.08 \pm 0.02$~M$_{\odot}$ \citep{Chaplin2013c}.
    
    For pulsating stars with a regular series of pressure modes in their frequency spectra, such as solar-type stars and red giant stars, common methods to identify $\Delta\nu$, and hence obtain the spherical harmonic geometry mode identification, are to directly fit the regular pattern in the frequency spectrum, or use a technique known as an {\'e}chelle diagram \citep{Chaplin2013c, Hekker2017a, Garcia_R_2019}. The former can be fairly computational expensive, whereas the latter can efficiently be used in either a manual or semi-automatic approach. An {\'e}chelle diagram is a frequency spectrum that has been divided into equal sections with a width that represents the distance between the rungs of a `ladder'. In so doing, for pressure modes that have the asymptotic property of being equally spaced in frequency, the spacing between a comb of frequencies can be derived and used to determine $\Delta\nu$. Since the definition of $\Delta\nu$ is the frequency spacing of pressure modes of consecutive radial order but the same angular degree and azimuthal order, the identification of $\Delta\nu$ means $\ell$ and $m$ values of pulsation modes can be assigned. An example of an {\'e}chelle diagram for the SLO star 16~Cyg~A is shown in Fig.~\ref{figure: 16 Cyg}, in which the dominant ridges corresponding to the $\ell = {0,1,2}$ and barely visible $\ell=3$ series of equally-spaced pressure modes are labelled. Strictly speaking, Eqn.~(\ref{equation: delta nu}) dictates that asymptotic pressure modes are exactly equally spaced in frequency. As can be seen in Fig.~\ref{figure: 16 Cyg}, however, this is not the case, and a small curvature in the {\'e}chelle ridges of pulsation modes is apparent. This discrepancy is caused by the physical conditions and processes within a star's envelope, such as the glitches in temperature and/or chemical composition profiles near partial ionisation zones \citep{Chaplin2013c, Hekker2017a, Garcia_R_2019}. These are details of the structure of a star that are not captured in Eqn.~(\ref{equation: delta nu}).
    
    \subsection{Gravity modes}
    \label{subsection: gravity}

    Gravity modes are standing waves with buoyancy (i.e., gravity) as their dominant restoring force, and as such they have lower frequencies compared to their pressure-mode counterparts. Gravity modes can only have non-radial ($\ell \geq 1$) spherical harmonic geometry and are defined by having frequencies lower than the Brunt-V{\"a}is{\"a}l{\"a} frequency, $N$, (also known as the buoyancy frequency) defined as:
    \begin{align}
        \label{equation: BV frequency}
        N^2 = g\left( \frac{1}{\gamma P} \frac{{\rm d}P}{{\rm d}r} - \frac{1}{\rho}\frac{{\rm d}\rho}{{\rm d}r} \right) ~ .
    \end{align}
    \noindent Analogous to the large frequency separation ($\Delta\nu$) for pressure modes, high-radial order gravity modes in the asymptotic regime are equally spaced in period and exhibit a characteristic period spacing, $\Pi_0$, which is defined as:
    \begin{align}
        \label{equation: asymptotic period}
        \Pi_0 = 2\pi^2 \left( \int_{r_1}^{r_2} N(r)\frac{{\rm d}r}{r}\right)^{-1} ~ ,
    \end{align}
    \noindent where the limits of the integral, $r_1$ and $r_2$, define the inner and outer radial coordinates of the gravity-mode pulsation cavity. The asymptotic period spacing, $\Pi_0$, which is also known as the buoyancy travel time \citep{Aerts2021a}, can also be calculated from fitting the periods of individual gravity modes, $P_{n \ell}$, via:
    \begin{align}
        \label{equation: gravity periods}
        P_{n \ell} = \frac{\Pi_0}{\sqrt{\ell(\ell+1)}} (|n| + \alpha) ~ ,
    \end{align}
    \noindent where $\alpha$ is phase term independent of a gravity mode's angular degree.

\begin{figure}
    \centering
    \includegraphics[width=0.8\textwidth]{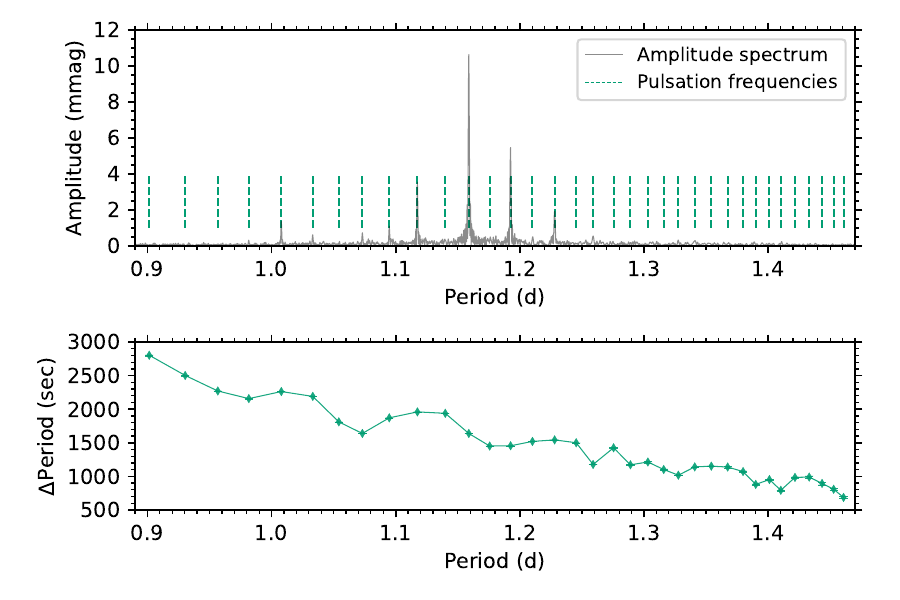}
    \caption{Top Panel: Amplitude spectrum as a function of period for the SPB star KIC~7760680 \citep{Bowman2021c}, in which the significant gravity-mode pulsation frequencies identified using iterative pre-whitening are marked. Bottom Panel: corresponding gravity-period spacing pattern of the prograde dipole (i.e. \{$\ell,m\} = \{1,1\}$) gravity-mode pulsation frequencies from the top panel. Note that error bars are plotted, but are generally smaller than the symbol size owing to long duration and extremely high-photometric precision of the {\it Kepler} space mission.}
    \label{figure: pattern}
    \end{figure}

    Given the asymptotic property of high-radial order gravity modes being equally spaced in period, a common diagnostic to analyse them and perform mode identification is to construct a gravity-mode period spacing diagram. This showcases the period differences of consecutive radial order ($n$) gravity modes of the same angular degree ($\ell$) and azimuthal order ($m$) plotted as a function of pulsation mode period. An example of a gravity-mode period spacing pattern is shown in Fig.~\ref{figure: pattern} for the slowly-pulsating B-type (SPB) star KIC~7760680. This star currently holds the record for the longest period spacing pattern in an SPB star, and thus has excellent precision on its interior physical processes (e.g. \citealt{Bowman2021c}). The definition of a gravity-mode period spacing pattern also means the hurdle of mode identification for pulsating stars with gravity modes is greatly alleviated.
   
    Once constructed, there are two main features of gravity-mode period spacing diagrams exploited by asteroseismology. First is that rotation induces a gradient in the pattern, as shown in the example of KIC~7760680 in Fig.~\ref{figure: pattern}, such that faster rotating stars have larger `tilts' in their patterns. Second, perturbations to a constant period spacing in addition to rotation come in the form of dips and `wiggles' caused by trapping of pulsation modes within zones containing chemical composition gradients. This is expected because of the chemical composition term in Eqns.~(\ref{equation: BV frequency}) and (\ref{equation: asymptotic period}), and can also be seen in the example of KIC~7760680 in Fig.~\ref{figure: pattern}. This means that the periods of individual gravity modes and the dips in gravity-mode period spacing patterns provide direct insight of the age of a star since the sizes of chemical gradient zones change during a star's evolution. However, there is a degeneracy among the mass, age and amount of chemical mixing when determining the morphology of the dips in a gravity-mode period-spacing pattern. For example, older and more-massive stars with a large amount of interior mixing can have similar gravity-mode period spacing patterns to younger less-massive stars with smaller amounts of interior mixing, which is then further compounded by an unknown rotation rate \citep{Bowman2020c}.

    It is worth noting that gravity modes are more commonly exploited in intermediate- and massive stars since gravity modes are evanescent in a convection zone. The amplitudes of gravity modes decay too quickly before reaching the surfaces of low-mass stars because of their thick convective envelopes. Whereas gravity modes have detectable surface amplitudes in main-sequence high- and intermediate-mass stars because of their radiative envelopes. Gravity modes are most sensitive to the region just outside of the hydrogen-burning convective core in such stars, with the core decreasing in size throughout the main sequence. The sensitivity of gravity modes to the size of a convective core is because the Brunt-V{\"a}s{\"a}il{\"a} frequency, $N$, is zero inside a convective core. Moreover, the chemical composition gradient zone just outside a convective core in addition to the energy transport mechanism at the convective-radiative interface causes a non-zero value of $N$ just outside a convective core. This so-called $N$-spike leads to trapping of the gravity modes, meaning they are most sensitive to the boundary layer between the convective core and radiative envelope. This is why the dips in gravity-mode period spacing patterns allow direct insight of the age of a star and the amount of mixing at this boundary.

    
    \subsection{Mixed pressure-gravity modes}
    \label{subsection: mixed modes}

    For main-sequence low-mass stars like the Sun, there is a large difference in the buoyancy and Lamb frequencies. However, as a star progresses through its main-sequence lifetime its buoyancy frequency increases and its Lamb frequency decreases. Eventually, once the star has entered the post-main sequence and become a red giant, these frequencies are close enough to allow mixed pressure-gravity modes, which are standing wave pulsations with gravity-mode characteristics whilst in the deep interior and pressure-mode characteristics near the surface. Mixed pressure-gravity modes are only expected for post-main sequence low- and intermediate-mass stars, and are commonly observed in red giant stars. Asteroseismology of mixed pressure-gravity modes in red giant stars has been hugely successful. Specifically, their period-spacing patterns have been used to distinguish the shell-hydrogen and core-helium burning phases of stellar evolution \citep{Bedding2011}. Also, since mixed pressure-gravity modes probe the full radius of a star, they can be used to probe the radial rotation profile (e.g. \citealt{Beck2012a}). Ensemble asteroseismic studies have revealed that tens of thousands of pulsating red giant stars have only small amounts of radial differential rotation \citep{Aerts2021a}. Additionally, precise masses and radii have been measured through the application of the asteroseismic scaling relations \citep{Chaplin2013c, Hekker2017a, Garcia_R_2019}. 
    
    At the other end of the HR~diagram, mixed pressure-gravity modes are already expected during the mid-to-late part of the main sequence phase of massive star evolution. This is because the buoyancy and Lamb frequencies of massive stars already become close to one another during the main sequence. Moreover, the frequencies of pressure and gravity modes in evolved stars are close enough to each other such that individual pulsation modes can undergo a from of mode interaction called avoided crossings, in which individual pulsation modes `bump' each other at specific epochs of stellar evolution \citep{ASTERO_BOOK}. Mixed pressure-gravity modes in massive stars can be exploited to provide precise constraints on the masses and ages of massive stars as well as their radial rotation profiles (e.g. \citealt{Burssens2023a}).

    
    \subsection{Pulsation excitation mechanisms}
    \label{subsection: mechanisms}

    There are two main pulsation excitation mechanisms. The first is the stochastic solar-like mechanism that operates within large convective envelopes and defines the pulsator group of SLOs \citep{Chaplin2013c}. The turbulent convection in the envelopes of main-sequence low-mass stars and evolved intermediate- and low-mass stars continuously drives and damps the resonant Eigenfrequencies of a star. An analogy would be taking a gong or bell into the middle of a sandstorm and hearing it vibrate because of the repeated impacts of many random sand grains. As discussed in Section~\ref{subsection: pressure}, the balance of driving and damping through the stochastic pulsation excitation mechanism caused by turbulent convection in the envelopes of low-mass stars typically produces moderate-to-high radial order pressure modes within a certain frequency range, which is centered on the frequency of maximum power: $\nu_{\rm max}$ \citep{Chaplin2013c}. Owing to the stochastic nature of this excitation mechanism, the excited pulsation modes appear as Lorentzian peaks in a frequency spectrum because they have finite mode lifetimes. This is illustrated for the SLO star 16~Cyg in Fig.~\ref{figure: 16 Cyg}.  

    The second main pulsation excitation mechanism is an opacity heat-engine mechanism in which mechanical work is converted from heat energy, like a piston in an engine. Within the radiative envelopes of massive and intermediate-mass stars, there are thin partial ionisation zones at specific temperatures for hydrogen, helium and metals such as iron and nickel, in which the medium exists in transition phase from being fully ionised deeper within the star and neutral closer to the surface. The presence of partially ionised metals within the hydrogen-rich envelope creates a local source of opacity. This increased opacity blocks the outward flow of radiation, which heats the zone and causes it to expand. After expanding, the radiation is able to flow through the layer unimpeded because the opacity has decreased, which means the source of heating has been removed so the layer cools down and contracts once again. The contraction of the layer leads to reforming the (partial) ionisation zone, thus recreating the source of opacity once again. This periodic expansion-contraction cycle allows heat from the star to be converted into mechanical work and excite pulsations \citep{Pamyat1999b}.

    
    \subsection{Other types of pulsation modes}  
    \label{subsection: other modes}

    In addition to pressure and gravity modes, there are other types of pulsation modes that are possible in different types of stars dependent on their properties. Similarly, these other types of pulsation modes are defined based on their dominant restoring forces. Mixed pressure-gravity modes are one such example and were discussed in Section~\ref{subsection: mixed modes}. Another example are gravito-inertial modes, which are commonly found in (rapidly) rotating intermediate- and high-mass stars (i.e. $\gamma$~Dor and SPB stars; see Sections~\ref{subsection: GDor} and \ref{subsection: SPB}), and have both rotation (i.e. the Coriolis force) and gravity as important restoring forces \citep{Aerts2021a}. Whereas in strongly magnetic pulsating stars, for example roAp stars (see Section~\ref{subsection: roAp}), magneto-acoustic modes are defined as such because both the Lorentz force and pressure gradient act as important restoring forces \citep{Kurtz1982c, Cunha2013}. Finally, in stars with strong enough magnetic fields, a pulsation mode may be suppressed or damped entirely, leading it to be converted into an Alfv{\'e}n wave \citep[e.g.][]{Fuller2015, Stello2016,Loi2017, Lecoanet2022a}.


\section{Data and methods of asteroseismology}  
\label{section: methods}

In this section, we provide an overview of the practical data analysis steps in asteroseismology. As an overview these include: (i) the observational component in which pulsation mode frequencies are extracted, optimised and identified; (ii) analysis of the frequencies to perform mode identification and inference of rotation or magnetic fields; and (iii) the modelling component in which the pulsation frequencies are exploited through a statistical comparison to numerical models of stellar structure and evolution to constrain their physical parameters. This full procedure is commonly referred to as forward asteroseismic modelling \citep{Bowman2020c, Aerts2021a}.

    
    \subsection{Observations of pulsation frequencies}
    \label{subsection: observations}

    Prior to mode identification of pulsation mode frequencies in terms of spherical harmonic geometry, one must first extract the frequencies themselves. This is achieved by assembling time-series data of a star's observable surface properties. Such time series data can be: (i) spectroscopic such that one quantifies periodic changes in the profiles of spectral lines, or (ii) time-series photometry, with the periodic variability of a star's integrated brightness as a function of time often called a light curve. Asteroseismology requires time series data to have a short cadence (i.e. short integration time) to avoid smearing of the pulsation signal and ensure the pulsation phases are well sampled. A time series should also be long in duration such that the frequency resolution, which is proportional to the inverse of the length of the data set, is sufficient to distinguish closely-spaced pulsation mode frequencies.

    To extract pulsation mode frequencies from a time series, it is common practice to employ Fourier analysis. Time series data are discrete and finite, so a frequency spectrum is calculated numerically using, for example, the discrete Fourier transform for unevenly sampled time series \citep{Kurtz1985b} or a Lomb-Scargle periodogram \citep{Scargle1982a}. The frequency spectrum is generally calculated up to the Nyquist frequency, $\nu_{\rm Nyquist}$, which is defined as:
    \begin{align}
        \label{equation: Nyquist}
        \nu_{\rm Nyquist} = \frac{1}{2 \Delta t} ~ ,
    \end{align}
    \noindent where $\Delta t$ is the cadence of the time series. The Nyquist frequency, $\nu_{\rm Nyquist}$, defines the upper frequency limit that is not under sampled, such that variability with a period of $1/\nu_{\rm Nyquist}$ is sampled with two epochs per period. Once the frequency spectrum has been calculated, peaks are denoted as statistically significant and represent pulsation mode frequencies based on satisfying a significance criterion. For example, this can be a false alarm probability (FAP) or a signal-to-noise ratio (SNR), in which the noise is taken to be the local average amplitude calculated in a specified frequency range at the location of the pulsation mode in the frequency spectrum (see e.g. \citealt{Kurtz2014, Bowman2021c}). 

    In the case of coherent pulsation modes often seen in intermediate- and high-mass stars, the pulsation mode frequencies typically appear as sinc functions with a full-width-half-maximum (FWHM) related to the Rayleigh frequency resolution of the input time series. In other words, the pulsation mode lifetimes can be considered infinite relative to the length of the time series data, because they are not resolved in time series data, even if it has a time span of a few decades. Once a significant coherent pulsation mode frequency has been identified, the optimal frequency, amplitude and phase can be determined by performing a least-squares fit of a multi-sinusoid function to the light curve using a model of the form
    \begin{align}
        \label{equation: sinusoid}
        \Delta\,m = \sum_{i}^{N} A_i\sin(2\pi\nu_{i}(t-t_0) + \phi_i) ~ ,
    \end{align}
    \noindent where $\Delta\,m$ is the model light curve with time stamps, $t$, and a pre-selected zero-point of the time series, $t_0$, and $A_i$ are the amplitudes, $\nu_i$ are the frequencies, $\phi_i$ are the phases of the $N$ significant frequencies included in the least-squares fit \citep{Bowman2021c}. In practice, this process for coherent pulsators is called iterative prewhitening and is employed in a recursive manner, such that a single pulsation mode is identified as significant, optimised by least squares fitting, included in the model (c.f. Eqn.~(\ref{equation: sinusoid})), and then the model is subtracted from the light curve to create a residual light curve. After which the process repeats again such that when a new frequency spectrum is calculated on the next iteration all significant pulsation mode frequencies that came before it have been removed. As an example, all of the significant gravity-mode pulsations in the period spacing pattern of the SPB star KIC~7760680 are labelled in the frequency spectrum in Fig.~\ref{figure: pattern}. This demonstrates that many of the gravity-mode frequencies are only marked as significant once the previous peaks have been removed during previous stages of iterative prewhitening --- see discussion and application by \citet{Bowman2021c}. 

    On the other hand, for stochastically excited pulsations, such as those in SLOs, the pulsation modes have finite lifetimes making them resolvable in time series of several months or longer. This means the peaks in the frequency spectra of SLOs are broad and well represented by a Lorentzian, as can be seen in Fig.~\ref{figure: 16 Cyg}. Therefore, the method of iterative prewhitening is not usually applied to SLOs since their stochastically excited pulsations require a very large number of sinusoids. Thus in the case of SLOs, the use of Lorentzian functions to directly fit the peaks in a frequency spectrum can be automated with great speed and efficiency if mode identification is first performed using an {\'e}chelle diagram to estimate $\Delta\nu$ a priori (e.g. \citealt{Corsaro2020a}).

    Once a list of all significant pulsation mode frequencies has been extracted via iterative prewhitening in the case of coherent pulsations or direct fitting using Lorentzians in the case of SLOs, the next step is to identify the spherical harmonic geometry of the modes (i.e. $\{n,\ell,m\}$). The best approach for this, however, depends on the type of pulsation mode, and the mass, evolutionary stage and pulsation excitation mechanism of the star. In general though, the methods for mode identification arise from applying the expectations from theory of how pulsation mode frequencies appear in a frequency spectrum such that numerical algorithms focused on pattern recognition have become increasingly more powerful and popular in the literature compared to visual inspection. The different methods of mode identification are generally specific to the type of pulsation modes, such as gravity modes, pressure modes, or mixed modes and how their asymptotic properties creating regularity in period and/or frequency, for example period spacing patterns, rotational multiplets, or {\'e}chelle diagrams, which were discussed in Section~\ref{section: pulsations}.


    \subsection{Asteroseismic scaling relations}
    \label{subsection: scaling relations}

    In the case of SLOs, once both $\Delta\nu$ and $\nu_{\rm max}$ (cf. Eqns.~\ref{equation: delta nu} and \ref{equation: nu-max}, respectively) have been identified, two simultaneous equations with two unknowns can be constructed, and rearranged to reveal the asteroseismic `scaling relations' for mass and radius:
    \begin{align}
        \label{equation: scaling relation mass}
        \frac{M}{{\rm M}_{\odot}} \simeq \left(\frac{\nu_{\rm max}}{\nu_{\rm max, \odot}}\right)^{3} \left(\frac{\Delta\nu}{\Delta\nu_{\odot}}\right)^{-4} \left(\frac{T_{\rm eff}}{T_{\rm eff,\odot}}\right)^{\frac{3}{2}} ~ ; \\
        \label{equation: scaling relation radius}
        \frac{R}{{\rm R}_{\odot}} \simeq \left(\frac{\nu_{\rm max}}{\nu_{\rm max, \odot}}\right) \left(\frac{\Delta\nu}{\Delta\nu_{\odot}}\right)^{-2} \left(\frac{T_{\rm eff}}{T_{\rm eff,\odot}}\right)^{\frac{1}{2}} ~ ,
    \end{align}
    \noindent where $\Delta\nu_{\odot} = 134.9$~$\muup$Hz is the large frequency separation of the Sun. These scaling relations are extremely valuable for a simple but precise estimate of fundamental parameters of solar-type and red giant stars, but assume that a star has the same structure as the Sun \citep{Hekker2020c}. The measurement of the surface gravity of the star through asteroseismology and the application of the scaling relations in Eqns.~(\ref{equation: scaling relation mass}) and (\ref{equation: scaling relation radius}) is generally more precise compared to estimates based on spectroscopy alone. This asteroseismic precision in turn leads to important constraints on the age of a star, which is challenging to infer directly, and yet highly relevant for many different fields of astrophysics \citep{Chaplin2013c, Hekker2017a, Garcia_R_2019}.
    
    However, it is important to note that the asteroseismic scaling relations in Eqns.~(\ref{equation: scaling relation mass}) and (\ref{equation: scaling relation radius}) do not include other stellar parameters, such as metallicity or other assumptions about the microphysics within a star, since they are normalised to observed properties of the Sun. Deviations from a constant frequency spacing in a frequency spectrum and {\'e}chelle diagram can be caused by rotation, or the presence of a chemical gradient and/or magnetic field within the pulsation cavity. Such effects may question the validity of the asteroseismic scaling relations for some stars, and a detailed asteroseismic study is then  needed to properly constrain the physics responsible. On the other hand, the application of the asteroseismic scaling relations to infer masses and radii of tens of thousands of red giant stars has been used to investigate new aspects of physics within stellar interiors, as well as aid galactic archaeology.


    \subsection{Measuring internal rotation}
    \label{subsection: rotation}
    
    \begin{figure}
    \centering
    \includegraphics[width=0.8\textwidth]{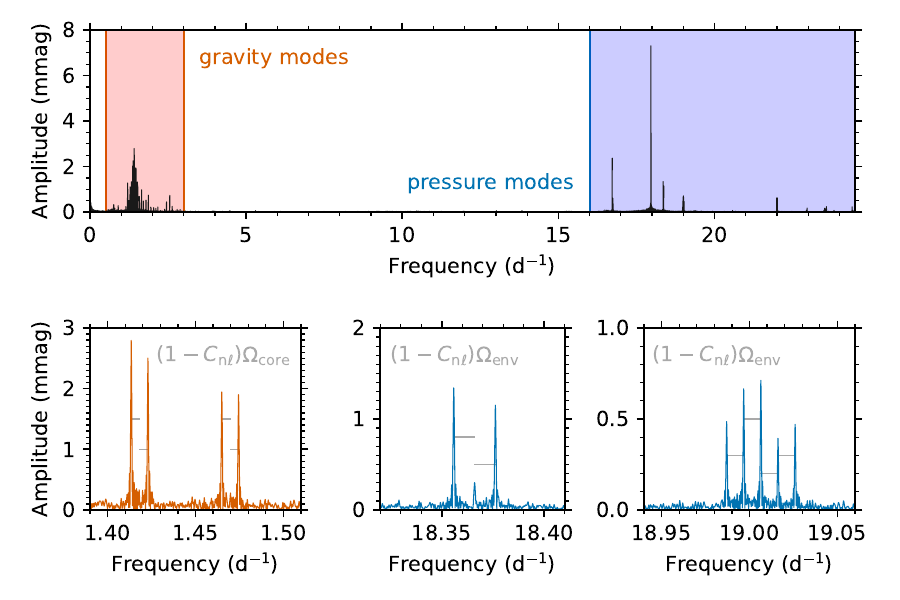}
    \caption{Top Panel: Frequency spectrum of the $\delta$~Scuti star KIC~11145123 \citep{Kurtz2014}, in which the frequency regimes of gravity and pressure modes are denoted in red and blue, respectively. The bottom panels show zoom-ins of two gravity-mode dipole triplets ($\ell=1$), a pressure-mode dipole triplet ($\ell=1$), and a pressure-mode quintuplet ($\ell=2$), from left-to-right. The frequency separation of components in a rotational multiplet reveals the rotation rate within the pulsation mode's cavity (cf. Eqn.~\ref{equation: Ledoux splitting}).}
    \label{figure: KIC1145123}
    \end{figure}

    \begin{figure}
    \centering
    \includegraphics[width=0.8\textwidth]{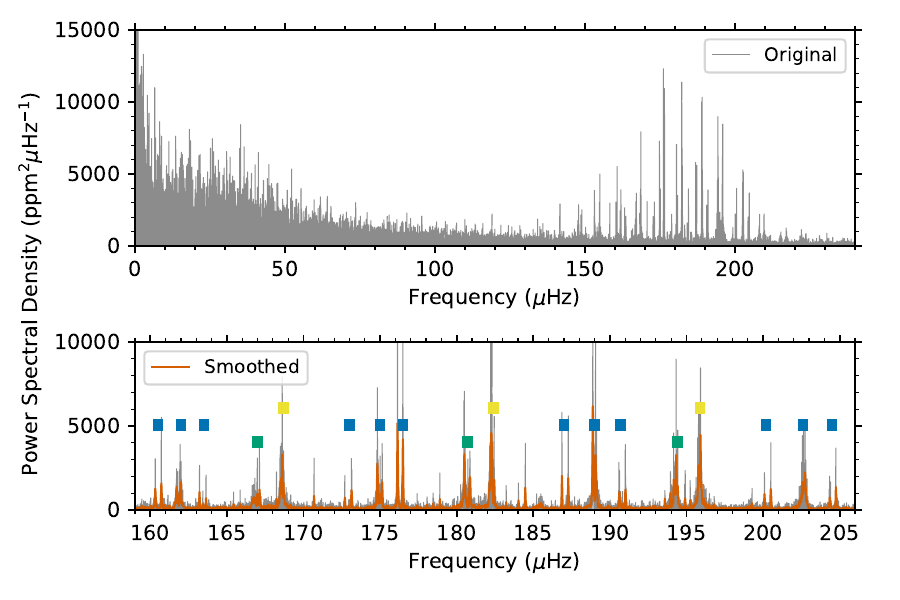}
    \caption{Top Panel: Frequency spectrum of the pulsating red giant star KIC~8366239 \citep{Beck2012a}, in which granulation from surface convection is dominant at low frequency, and the regular series of pulsation frequencies at high frequency can be seen. Bottom Panel: zoom-in of the pulsation frequencies, with radial ($\ell=0$), rotationally-split dipole ($\ell=1$) and quadrupole ($\ell=2$) modes marked by yellow, blue and green squares, respectively. The dipole modes are split by rotation into rotational multiplets.}
    \label{figure: KIC8366239}
    \end{figure}

    Since all stars rotate to some extent, the Coriolis force is important to consider for two main reasons. First, rotation breaks the assumption of spherical symmetry for stellar structure, because it `flattens' a star making its equatorial radius larger than its polar radius. This leads to gravity darkening, such that the equator has a lower effective temperature than the pole \citep{Kippenhahn_BOOK, Langer2012}. Second, rotation has a significant perturbative effect to pulsation mode frequencies through the Doppler shift. When combined, these two effects can drastically impact the excitation mechanisms of a pulsating star and the frequencies that are excited \citep{Townsend2005e, Szewczuk2017a}. For example, prograde (defined hereafter as $m>0$) modes as those that travel in the direction of a star's rotation are imparted a positive Doppler shift to higher frequency, whereas retrograde (defined hereafter as $m<0$) modes experience a Doppler shift to lower frequencies in the corotating reference frame of the star. Note that the observer is located in the inertial reference frame, meaning that the positive and negative frequency shifts for prograde and retrograde modes are mirrored when using the same azimuthal-order sign convention because of the change of reference frame \citep{ASTERO_BOOK}.

    The simplest scenario is a star that is slowly rotating as a rigid body with a rotation rate of $\Omega$, such that its radial rotation profile is constant. Slowly rotating here is defined as rotating at less than about 15\% of its critical breakup velocity \citep{Aerts2021a}. In such cases, a non-radial pulsation mode frequency, $\omega_{n, \ell, m}$, is given by:
    \begin{align}
        \label{equation: Ledoux splitting}
        \omega_{n, \ell, m} = \omega_{n, \ell} + m(1-C_{n, \ell})\Omega ~ ,
    \end{align}
    \noindent where $C_{n, \ell}$ is the Ledoux constant, and $\omega_{n, \ell}$, is the degenerate (i.e. $m=0$) pulsation mode frequency solution in the non-rotating scenario. The Ledoux constant takes a value of $0 \leq C_{n, \ell} \leq 1$ and is dependent on the pulsation mode geometry and stellar structure (e.g. \citealt{Burssens2023a}). However, a first-order approximation is that dipole pressure modes have $C_{n, \ell=1} \simeq 0$, and dipole gravity modes have $C_{n, \ell=1} \simeq 0.5$ \citep{ASTERO_BOOK}. Equation~(\ref{equation: Ledoux splitting}) gives rise to symmetric rotational multiplets in the frequency spectrum of a slowly-rotating rigid-body star by lifting the degeneracy of pulsations modes of the same $n$ and $\ell$, but different $m$. Moreover, because of the quantum number selection rule of $-\ell \leq m \leq +\ell$, the number of component frequencies in a multiplet reveals the $\ell$ and corresponding $m$ values. For example, a triplet corresponds to dipole (i.e. $\ell=1$) pulsation modes separated into $m \in \{-1,0,+1\}$, whereas a quintuplet corresponds to quadrupole (i.e. $\ell=2$) pulsation modes separated into $m \in \{-2,-1,0,+1,+2\}$, and so on. However, the inclination of the star has a big impact on the relative amplitudes of the different components within a multiplet \citep{ASTERO_BOOK}.

    Rotational multiplets can exist for non-radial pressure and gravity modes, and are symmetric as long as the rotation rate is slow enough for the first-order approximation to apply (c.f. Eqn.~\ref{equation: Ledoux splitting}), making them an effective method of mode identification (e.g. \citealt{Aerts2003d, Kurtz2014}). For more rapidly stars, or stars with non-rigid radial rotation profiles, the first-order Ledoux approximation in Eqn.~(\ref{equation: Ledoux splitting}) is no longer valid and an higher-order formalism is required \citep{Dziembowski1992c, Suarez2010b}. Examples of slowly rotating pulsating stars with rotational multiplets are shown in Figs.~\ref{figure: KIC1145123} and \ref{figure: KIC8366239}. The former example is the main-sequence $\delta$~Scuti star KIC~11145123 studied by \citet{Kurtz2014}, and was discovered to have a near-rigid radial rotation profile with a period of approximately 100~d. The latter example is the red giant star KIC~8366239 \citep{Beck2012a} which exhibits rotational splitting of approximately 0.3~$\muup$Hz in its dipole mixed modes, as can be seen in its frequency spectrum in Fig.~\ref{figure: KIC8366239}, which corresponds to a core-rotation period of about 40~d.

    In addition to rotational multiplets, gravity-mode period spacing patterns are powerful diagnostic diagrams in gravity-mode asteroseismology for measuring interior rotation rates of pulsating stars, as introduced in Section~\ref{subsection: gravity}. This is because high-radial order gravity modes of consecutive radial order ($n$), but the same angular degree ($\ell$) and azimuthal order ($m$), are equally spaced in period when plotted as a function of their pulsation period for a non-rotating and chemically homogeneous star. Rotation is a strong perturbation to pulsation periods through the Doppler shift, meaning that the Coriolis force induces a gradient in a gravity-mode period spacing pattern, with faster rotating stars having larger gradients \citep{Bouabid2013}. Owing to the necessity of defining the observer to be in the inertial reference frame, prograde pulsation modes (i.e. $m>0$) have a negative gradient in a gravity-mode period spacing pattern. Whereas a retrograde (i.e. $m<0$) period spacing pattern has positive gradient. Finally, the gradient of a zonal (i.e. $m=0$) gravity-mode period spacing pattern is approximately zero, except for the fastest of rotators \citep{Bouabid2013}. Ensemble studies have revealed that prograde dipole (i.e. $\{\ell,m\} = \{1,1\}$) gravity-mode period spacing patterns are by far the most common spherical harmonics geometry for pulsations in intermediate- and high-mass stars \citep{Aerts2021a}.

    \begin{figure}
    \centering
    \includegraphics[width=0.4\textwidth]{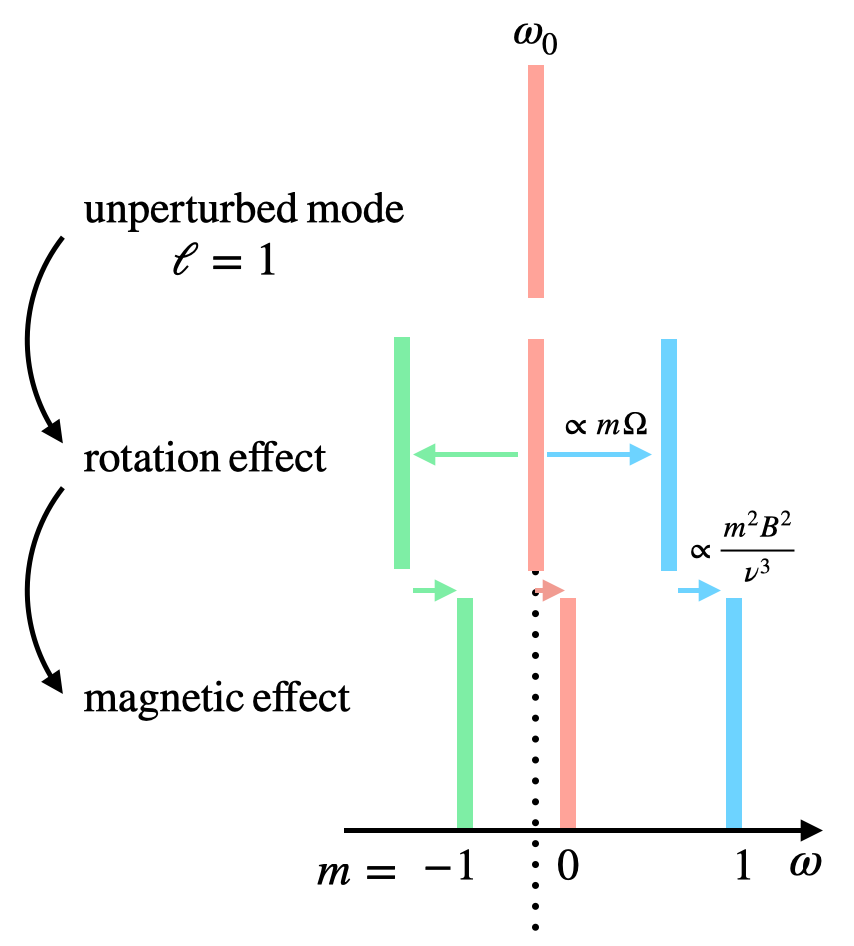}
    \caption{Schematic effect of rotation and simple magnetic fields (at first order) on an gravity-mode frequency, lifting the degeneracy of the azimuthal order $m$ of the mode and shifting the components to higher frequencies. Adapted from \citet{Bugnet2021a}.}
    \label{figure: magnetic}
    \end{figure}

     
    \subsection{Measuring internal magnetism}
    \label{subsection: magnetism}
    
    Magnetic fields also impact the structure of stars as well as the frequency of pulsations through the Lorentz force. Fortunately, the strength and geometry of interior magnetic fields can be probed with asteroseismology. In main-sequence stars with radiative envelopes, large-scale magnetic fields can significantly perturb gravity-mode pulsation frequencies \citep{Hasan2005}. For red giant stars, mixed pressure-gravity modes can be used to probe magnetic fields within their radiative cores --- see predictions by \citet{Loi_S_2020c} and \citet{Bugnet2021a}. This is achieved by modelling the asymmetry induced by magnetic fields in the asteroseismic multiplets in a frequency spectrum. Indeed the frequency of mixed modes increases when the stellar interior is magnetized, and at first order gravity mode frequencies are perturbed following:

    \begin{equation}
        \omega_{n,\ell,m} \approx \omega_{n, \ell} + \frac{m^2 B^2}{\omega_{n, \ell}^3} \mathcal{R}_{n,\ell} ~ ,
        \label{eq:mag}
    \end{equation}

    \noindent where $B$ is the strength of the radial component of the magnetic field at the hydrogen-burning shell, and $\mathcal{R}_{n,\ell}$ is a factor that depends on the parameters of the star and the pulsation mode, akin to the Ledoux constant for rotation. Because of the $m^2$ dependency of the magnetic effect in Eqn.~(\ref{eq:mag}), the mixed mode pattern, which is symmetric in the case of a slowly-rotating non-magnetized star, becomes asymmetric as shown in Fig.~\ref{figure: magnetic}. The impact of the magnetic perturbation effect is most important for lower frequency gravity modes due its dependency on the term $1/\omega_{n,\ell}$ in Eqn.~(\ref{eq:mag}).

    Internal magnetic fields in red giant stars were detected for the first time by \cite{Li_G_2022a} through the study of gravity-dominated mixed modes. This confirms the presence of strong magnetic fields in the cores of evolved low- and intermediate-mass stars. Previously this was postulated based on the observed suppressed amplitudes of mixed modes in such stars \citep{Fuller2015, Stello2016}. Magneto-asteroseismology has now become a very active field of research, aiming at evaluating the impact of magnetism on the internal dynamics and evolution of main-sequence stars (e.g. \citealt{Lecoanet2022a}).

    
    \subsection{Forward asteroseismic modelling} 
    \label{subsection: modelling}

    The quantitative comparison of observed pulsation frequencies to those predicted from a multi-dimensional grid of theoretical stellar structure models is called forward asteroseismic modelling \citep{Aerts2018b, Aerts2021a}. In the typical forward asteroseismic modelling scenario, a grid of stellar models is computed for a particular type of pulsator or group of similar pulsators, since a dense grid of models covering the entire HR~diagram is not feasible. These numerical models include a set of assumptions for the microphysics, such as opacities, chemical mixture, and equation of state. The structure models contain free parameters to be determined, such as mass, initial hydrogen and metal mass fractions (i.e. $X$ and $Z$), age, but also include prescriptions for unconstrained physical processes such as chemical mixing and rotation \citep{Bowman2020c}. A stellar structure model grid is calculated to cover the required parameter space inferred from spectroscopic estimates of effective temperature and luminosity, or surface gravity as a proxy, in the HR diagram, whilst varying the chosen free parameters. 
    
    The corresponding theoretical pulsation frequencies for each structure model in the grid are obtained by numerically perturbing the equilibrium structure models to calculate the Eigenfrequencies (e.g. \citealt{Townsend2018a}) for a range of spherical harmonic mode geometries. The parameter range to calculate is informed by observations, such as frequencies from frequency spectra and mode identification from rotational multiplets and/or gravity-mode period spacing patterns. Therefore, forward asteroseismic modelling is a model selection problem requiring a statistical framework \citep{Aerts2018b}, such that the parameters of the best-fitting structure model are those that most accurately represent the physics of the observed star. 
    
    It is worth noting that there are other approaches in asteroseismic modelling to constrain the interior physics of pulsating stars. For example, rotational and structural inversions (e.g. \citealt{Bellinger2017b, Vanlaer2023a}). There is also growing interest in the application of machine-learning techniques for efficient ensemble asteroseismic modelling (e.g. \citealt{Mombarg2021a}).


\section{Highlights of progress in understanding different types of pulsating stars}
\label{section: results}

In this section, we present an overview of important (and non-exhaustive) highlights of asteroseismology results in the last couple of decades. The majority of these results are primarily thanks to long-term and high-precision light curves from space telescopes that provided the first continuous data sets spanning several years, such as the {\it Kepler} (2009-2014; \citealt{Borucki2010}) and ongoing TESS \citep{Ricker2015} space missions. 

\begin{figure}
\centering
\includegraphics[width=0.95\textwidth]{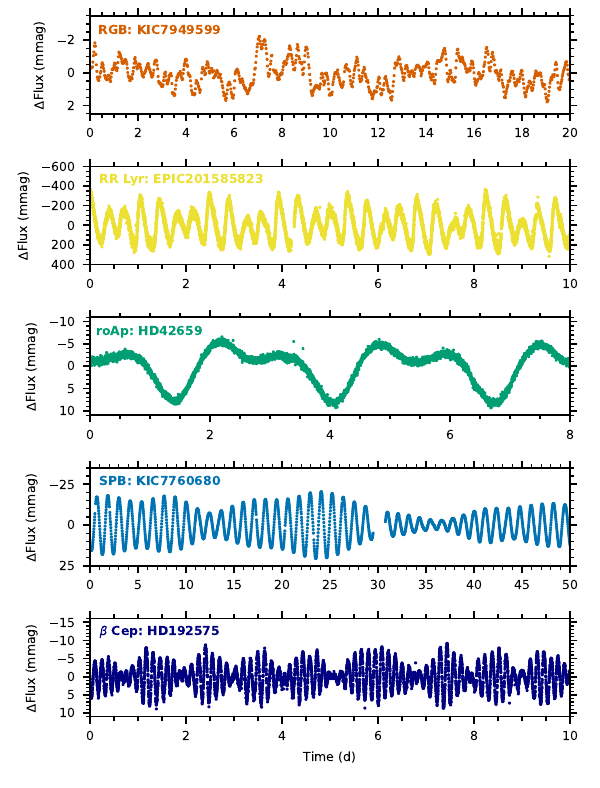}
\caption{Example light curves of a selection of different types of pulsating stars discussed in Section~\ref{section: results}. Note that the ordinate and abscissa scales are not the same because of the different peak-to-peak brightness and timescales caused by pulsations among the selected pulsator classes.}
\label{figure: light curves}
\end{figure}

Asteroseismology continues to expand to include a broad range of masses and evolutionary stages. New space photometry providing longer light curves allows higher frequency precision and allows us to probe different parameter spaces. An asteroseismic HR~diagram is shown in Fig.~\ref{figure: HRD}, which contains effectively all of the known types of pulsating stars --- see \citet{Kurtz2022a}. However, it is not possible to provide an exhaustive discussion of all pulsator types in this introductory chapter on Asteroseismology. Instead, we focus on the historic and most common pulsator types. The approximate ranges in effective temperature and luminosity, typical pulsation periods and amplitudes for the various pulsator types are provided in Table~\ref{table: pulsator classes}. A selection of example light curves are shown in Fig.~\ref{figure: light curves}, which demonstrate the diversity in amplitudes and time scales of pulsating stars across the HR~diagram.

\begin{table}
\TBL{\caption{Properties of a non-exhaustive list of different pulsator classes, updated from the original version by \citet{ASTERO_BOOK}. Columns include the name, which is sometimes a written as a commonly used acronym or label in the literature, mode type (where p=pressure, g=gravity and S=strange\footnotemark{a}, and where ? denotes uncertainty in the mode type), typical periods and photometric amplitudes, and the approximate parameter space in the HR~diagram defined by ranges in effective temperature ($T_{\rm eff}$) and luminosity ($L$), which are also shown in Fig.~\ref{figure: HRD}.}
\label{table: pulsator classes}}
{\begin{tabular*}{\textwidth}{@{\extracolsep{\fill}}@{}ccccccc@{}}
\toprule
\multicolumn{1}{@{}c}{\TCH{Pulsator}} &
\multicolumn{1}{c}{\TCH{Mode}} &
\multicolumn{1}{c}{\TCH{Pulsation}} &
\multicolumn{1}{c}{\TCH{Pulsation}} &
\multicolumn{1}{c}{\TCH{$T_{\rm eff}$}} &
\multicolumn{1}{c}{\TCH{$L$}} \\
\multicolumn{1}{@{}c}{\TCH{class name}} &
\multicolumn{1}{c}{\TCH{type}} &
\multicolumn{1}{c}{\TCH{periods}} &
\multicolumn{1}{c}{\TCH{amplitudes\footnotemark{b}\footnotemark{c}}} &
\multicolumn{1}{c}{\TCH{\rm (K})} &
\multicolumn{1}{c}{\TCH{\rm (L$_{\rm \odot}$)}} \\
\colrule
\multicolumn{6}{l}{\TCH{Main Sequence}} \\
\hline
Solar-type (SLO) & p & $1-10$~min & $<10$~ppm & $5000-6600$ & $0.3-10$ \\
$\gamma$~Dor & g & $0.3-5$~d & $<50$~mmag & $6500-8500$ & $2-20$ \\
$\delta$~Sct & p & $0.01-0.25$~d & $<0.3$~mag & $6400-9500$ & $3-100$ \\
$\delta$~Sct & g & $0.25-5$~d & $<1$~mmag & $6400-9500$ & $3-100$ \\
roAp & p & $4-24$~min & $<10$~mmag & $6300-8500$ & $5-40$ \\
SPB & g & $0.5-5$~d & $<50$~mmag & $11000-20000$ & $10^2-10^4$ \\
Be & g & $0.1-5$~d & $<50$~mmag & $11000-30000$ & $10^2-10^4$ \\
$\beta$~Cep & p,g & $0.05-3$~d & $<50$~mmag & $15000-30000$ & $10^3-10^4$ \\
SLF & g & $0.01-10$~d & $<50$~mmag & $10000-50000$ & $10^2-10^6$ \\
\hline
\multicolumn{6}{l}{\TCH{Pre-Main Sequence}} \\
\hline
T~Tauri & p,g & $0.05-5$~d & $<5$~mmag &  &  \\
Herbig Ae/Be & p & $1-8$~hr & $<5$~mmag &  &  \\
\hline
\multicolumn{6}{l}{\TCH{Post-Main Sequence}} \\
\hline
Subgiant (SLO) & p,g & $15-30$~min & $<50$~ppm & $4800-6300$ & $1-10$  \\
Red Giant (SLO) & p,g & $1-12$~hr & $<100$~ppm & $4500-5000$ & $0.3-100$  \\
Type~{\sc I} Cepheid & p & $1-100$~d & $<1$~mag & $3500-7000$ & $10^2-10^{5.5}$ \\
Type~{\sc II} Cepheid (BL~Her) & p & $1-5$~d & $<1$~mag & $5000-8000$ & $10^2-10^4$ \\
Type~{\sc II} Cepheid (W~Vir) & p & $10-20$~d & $<1$~mag & $5000-8000$ & $10^2-10^4$ \\
Type~{\sc II} Cepheid (RV~Tauri) & p & $20-150$~d & $<3$~mag & $4000-8000$ & $10^{3.2}-10^{4.2}$ \\

RR~Lyr & p & $0.3-0.5$~d & $<1.5$~mag & $6000-8000$ & $25-50$ \\
Semi-regular (SRa, SRb, SRc) & p & $>80$~d & $<8$~mag & $2800-5600$ & $10^{1.5}-10^{2.5}$ \\
Semi-regular (SRd) & p & $<80$~d & $<1$~mag & $2800-5600$ & $10^{1.5}-10^{2.5}$ \\
Mira & p & $100-1000$~d & $<10$~mag & $2500-3500$ & $10^{2.5}-10^{4.0}$ \\
$\alpha$~Cyg & g,S? & $10-100$~d & $<0.3$~mag & $2800-10000$ & $10^{2.5}-10^{4.5}$ \\
LBV (S~Dor) & g,S & $2-40$~d & $<0.1$~mag & $6300-16000$ & $10^{5.5}-10^{6.5}$ \\
sdBV & p & $80-800$~sec & $<0.1$~mag & $15000-30000$ & $15-150$ \\
sdBV & g & $0.5-3$~hr  & $<10$~mmag & $25000-40000$ & $15-400$ \\
GW~Vir & g & $5-80$~min & $<0.2$~mag & $60000-120000$ & $30-3000$ \\
V777~Her & g & $2-16$~min & $<0.2$~mag & $25000-40000$ & $0.1-5$ \\
ZZ~Ceti & g & $0.5-25$~min & $<0.2$~mag & $9000-14000$ & $10^{-2.6}-10^{-2.2}$ \\
R~CrB & g? & $40-100$~d & $<50$~mmag & $3100-16000$ & $10^{3.5}-10^{4.5}$ \\
PV~Tel & S & $\sim20$~d & $<50$~mmag & $3100-16000$ & $10^{3.5}-10^{4.5}$ \\
V652~Her & g,S? & $\sim0.1$~d & $<50$~mmag & $3100-16000$ & $10^{3.5}-10^{4.5}$ \\




\botrule
\end{tabular*}}{%
\begin{tablenotes}
\footnotetext[a]{Strange modes are theoretically predicted for massive stars near the Eddington limit (i.e. large $L/M$ ratios) and are non-linear radial instabilities caused by the strong enhancements in the non-adiabatic partial ionization zones of helium and heavy elements, resulting in long period and high-amplitude outburst-like pulsations \citep{Glatzel1999b, ASTERO_BOOK}.}
\footnotetext[b]{The units here are photometric pulsation amplitudes and are expressed in the most common form used in the literature for each pulsator class, with an approximate conversion being: 1~$\muup{\rm mag} \simeq 1$~ppm.}
\footnotetext[c]{Pulsation amplitudes are those typically observed in broadband optical photometry, but in reality these depend on the colour of the star and the wavelength passband of the instrument. Note also that spectroscopic amplitudes of pulsations can differ to their photometric counterparts although they are generally less available in the literature.}
\end{tablenotes}
}%
\end{table}

 
    \subsection{Solar-type stars}
    \label{subsection: SLO}

    Solar-type stars are main-sequence stars with a mass similar to that of the Sun ($0.8 \lesssim M \lesssim 1.2$~M$_\odot$). Owing to their similar structures they therefore also pulsate like the Sun, with stochastically excited pressure modes driven by turbulent convection in their convective envelopes \citep{Chaplin2013c}. Unfortunately, solar-type stars are typically quite faint in the Universe and their pulsation amplitudes are very small (i.e. of order a few ppm), making them challenging to detect and analyse. An excellent example of a pulsating solar-type star with a frequency spectra similar to the Sun is 16~Cyg~A, which is shown in Fig.~\ref{figure: 16 Cyg}. In the specific application of seismology to the Sun -- Helioseismology -- a large fraction of the internal density and rotation profiles, as well as the depth of the convective envelope has been measured with great accuracy \citep{Christensen1991}.


    \subsection{Red giant stars}
    \label{subsection: red giants}

    Red giant stars represent a future evolutionary stage of our Sun. Once the hydrogen is exhausted in the core at the end of main sequence, an intermediate- or low-mass star expands and eventually becomes a red giant star. Stochastically excited pulsations are commonly observed at the surfaces of red giant stars, making them SLOs. An example light curve of the red giant star KIC~7949599 \citep{Beck2012a} is shown in Fig.~\ref{figure: light curves}. As their structures change as they ascend the red giant branch in the HR~diagram, the pulsations become more strongly coupled with gravity-mode pulsations in the deep radiative interior. As a result, it is typical to observe mixed pressure-gravity modes for red giant stars. Red clump stars are red giant stars specifically in the core-helium burning phase (i.e. horizontal branch stars), which leads to their pulsations having much stronger mixed-mode signatures. Red giant stars are broadly studied in asteroseismology, as the analysis of pulsation modes provide tight constraints on masses and radii via the asteroseismic scaling relations (cf. Eqn.~(\ref{equation: scaling relation mass}) and (\ref{equation: scaling relation radius})) --- see reviews by \citealt{Chaplin2013c} and \citet{Hekker2017a} and \citet{Garcia_R_2019}. Moreover, the mixed modes of red giants stars directly probe the physical processes of their deep interiors, which pressure modes alone in solar-type stars cannot achieve \citep{Christensen1991}. Recent breakthroughs include statistics on the core rotation rates of thousands of red giant stars indicating that a strong angular momentum transport must be operating in the deep interior (e.g. \citealt{Gehan2018}), as well as the presence of strong magnetic fields in the cores of red giant stars \citep{Li_G_2022a}.

    
    \subsection{$\gamma$~Dor stars}
    \label{subsection: GDor}

    The $\gamma$~Doradus ($\gamma$~Dor) stars are main-sequence pulsators with spectral types between late-A and early-F, hence masses between about $1.5 \leq M \leq 2.5$~M$_{\odot}$. The high-radial order gravity modes in $\gamma$~Dor stars are excited by a flux-blocking mechanism: the convective turnover timescale at the base of their thin surface convective envelopes is comparable to the pulsation periods of gravity modes, allowing a modulation of the outward flowing radiation sufficient to excite gravity modes \citep{Guzik2000a, Dupret2004a}. This is similar to the opacity-driven heat engine mechanism, except that it is the convective motions rather than opacity acting as the `value' in the heat engine of $\gamma$~Dor stars. The gravity-mode period spacing patterns of $\gamma$~Dor stars typically span a large range of radial orders and have been exploited to measure the near-core rotation rates for hundreds of stars \citep{VanReeth2016a, Li_G_2020a}. Forward asteroseismic modelling of gravity modes in ensembles of $\gamma$~Dor stars has provided precise and accurate masses and ages, and demonstrated generally that the masses of the hydrogen-burning convective cores in such stars are underestimated by stellar evolution models, so that additional mixing mechanisms are needed specifically at the core-envelope boundary layer \citep{Mombarg2019a}. Moreover, the near-rigid radial rotation profiles in their envelopes and on average fast overall rotation rates of hundreds of $\gamma$~Dor stars are currently a challenge to explain with current angular momentum transport theory \citep{Ouazzani2019a}.


    \subsection{$\delta$~Sct stars}
    \label{subsection: DSct}
        
    The $\delta$~Scuti ($\delta$~Sct) stars have spectral types of A and early-F, and masses between about $1.5 \leq M \leq 3$~M$_{\odot}$. They lie at the intersection of the main-sequence and the classical Cepheid instability strip in the HR~diagram, and span the pre-main sequence (e.g. \citealt{Zwintz2014b}) to the post-main sequence \citep{Breger2000b}. The heat-engine mechanism operating in the helium partial ionisation zone at 50\,000~K is efficient in exciting low-radial order pressure modes with periods spanning from tens of minutes up to several hours \citep{Breger2000b}. Some $\delta$~Sct stars are also observed to have gravity-mode pulsations with periods between a few days and several hours making them hybrid pulsators (e.g. \citet{Uytterhoeven2011}), which models struggle to explain. The small fraction of $\delta$~Sct stars with higher-radial-order pressure modes with periods of order tens of minutes are thought to be excited by turbulent pressure \citep{Antoci2019a}. Given their often hybrid nature, the use of rotational multiplets in slowly rotating $\delta$~Sct stars allows the radial rotation profile to be extracted. \citet{Kurtz2014} demonstrated how KIC~11145123 has a nearly uniform interior rotation period of approximately 100~d, which is shown in Fig.~\ref{figure: KIC1145123}. However, $\delta$~Sct stars are often rapid rotators which causes their frequency spectra to be quite difficult to interpret and forward asteroseismic modelling challenging \citep{Daszy2021a}. Some $\delta$~Sct stars show regular patterns in their frequency spectra (e.g. \citealt{Bedding2020a}), allowing the application of {\'e}chelle diagrams for mode identification and forward asteroseismic modelling (e.g. \citealt{Murphy_S_2023a}).


    \subsection{roAp stars}
    \label{subsection: roAp}
        
    The rapidly oscillating Ap (roAp) stars are a sub-group of chemically-peculiar pulsating A-type stars located near the main sequence within the classical instability strip \citep{Cunha2002d}. They have strong, predominantly dipolar magnetic fields, slow rotation rates, and high-radial order and low-angular degree magneto-pressure pulsation modes with periods of order several minutes \citep{Kurtz1990a}. The exact pulsation excitation mechanism(s) at work in roAp stars is not fully understood, but it is likely a combination of the heat-engine mechanism operating in the hydrogen partial ionisation zone and turbulent pressure \citep{Cunha2013}. The pulsation axis in roAp stars is misaligned with respect to both the rotation and magnetic axes, giving rise to the oblique pulsator model \citep{Kurtz1982c}. This allows the pulsations to be viewed from different aspects as the star rotates, giving rise to rotational multiplets in their frequency spectra. An example of a light curve of the roAp star HD~42659 \citep{Holdsworth2021b} is shown in Fig.~\ref{figure: light curves}. Only about 100 roAp stars are known making them a particularly rare type of pulsating star \citep{Holdsworth2024a}.


    \subsection{RR Lyrae stars}
    \label{subsection: RR Lyr}

    RR Lyrae (RR Lyr) stars are often referred to as classical pulsators because of their dominant high-amplitude radial-mode pulsations caused by the heat-engine mechanism, and they have been used historically as standard candles in astrophysics. They are metal-poor and old (i.e. Population~{\sc II}) giant stars with masses smaller than about 1~M$_{\odot}$ on the horizontal branch \citep{ASTERO_BOOK}. RR~Lyr stars are subdivided into RRab, RRc and RRd stars. The former are classified by having highly non-sinusoidal light curves caused predominantly by non-linear behaviour in the fundamental radial pulsation mode. RRc stars have sinusoidal light curves dominated by the first-overtone radial mode \citep{Stellingwerf1975a}, whereas RRd stars are double-mode RR Lyr stars with both the fundamental and first overtone radial modes simultaneously excited. An example light curve of an RRd star, EPIC~201585823 \citep{Kurtz2016a}, is shown in Fig.~\ref{figure: light curves}. Some RR~Lyr stars are not perfectly periodic and exhibit a form of amplitude and phase modulation called the Tseraskaya-Blazhko effect \citep{Blazhko1907, Kurtz2022a}. Another phenomenon found in RR~Lyr stars is period doubling which is caused by half-integer resonances (e.g. 9:2) among the overtone and fundamental radial modes \citep{Kollath2011}. In a light curve, period doubling appears as alternating minima and maxima in the brightness excursions \citep{Szabo2010}.


    \subsection{Cepheid variables}
    \label{subsection: Cepheids}

    Cepheid variables are also known as classical pulsators from their heat-engine radial-mode pulsations and from being standard candles. They are evolved horizontal branch stars crossing the classical instability strip in the HR~diagram \citep{ASTERO_BOOK}. Cepheids are subdivided into two groups. Type~{\sc I} Cepheids are Population~{\sc I} stars typically crossing the classical instability strip for the first time, and have radial-mode pulsations with periods between 1 and 100~d. The longest-period type~{\sc I} Cepheids are also the brightest and referred to as ultra-luminous Cepheids. The period ratios of first-overtone and fundamental radial modes elucidate their evolutionary age \citep{Petersen1973}. Type~{\sc II} Cepheids are Population~{\sc II} stars transitioning from the horizontal branch to the asymptotic giant branch \citep{Wallerstein2002}. They also pulsate in radial modes, but are subdivided into three classes because they have heterogeneous pulsation periods: BL Herculis stars have periods between 1 and 5 d; W Virginis stars have periods between 10 and 20 d; and RV Tauri stars have periods longer than 20 d \citep{ASTERO_BOOK}. The pulsations in Cepheids are generally low-radial order, such as the fundamental, first overtone or second overtone radial modes, or a combination of these.

 
    \subsection{SPB and Be stars}
    \label{subsection: SPB}
        
    The slowly pulsating B-type (SPB) stars are dwarf stars with spectral types between B3 and B9, and have high-radial order gravity modes excited by the heat-engine mechanism operating in the iron-bump at 200\,000~K \citep{Dziembowski1993e, Pamyat1999b}. Forward asteroseismic modelling of their gravity-mode period spacing patterns has revealed their near-core rotation rates \citep{Papics2017a, Szewczuk2021a} as well as a diversity in the amount of interior mixing \citep{Pedersen2021a}. The light curve of the SPB star KIC~7760680 \citep{Bowman2021c} is shown in Fig.~\ref{figure: light curves}, and its frequency spectrum and gravity-mode period spacing pattern are shown in Fig.~\ref{figure: pattern}. Interestingly, a handful of SPB stars host a strong large-scale magnetic field at their surfaces (e.g. \citealt{Buysschaert2018c}). Through a combination of spectropolarimetry, asteroseismology and MHD simulations, this has allowed magneto-asteroseismology to probe the strength of the magnetic field in the deep interior near the convective core in the SPB star HD~43317 \citep{Lecoanet2022a}. The analysis suggests the presence of a core-dynamo strengthening the large-scale fossil field. Finally, about 20\% of dwarf B-type stars rotate sufficiently rapidly to create a so-called decretion disk of ejected circumstellar material and are classified as Be stars \citep{Rivinius2013c}. The observational signatures of rapid rotation and the presence of a decretion disk are made based on observing strong emission lines in spectroscopy, and are also known to be time-dependent. The Be stars are commonly found to pulsate, with rotation rate and pulsator fraction being correlated \citep{Labadie-Bartz2022a}. The rapid rotation and pulsations provide very efficient angular momentum transport and chemical mixing \citep{Huat2009c, Neiner2020b}.

    
    \subsection{$\beta$~Cep stars}
    \label{subsection: BCep}

    The $\beta$~Cephei ($\beta$~Cep) stars are massive stars with spectral types typically ranging between late-O and early-B whilst on the main sequence, thus they have birth masses between approximately 8 and 30~M$_{\odot}$ \citep{Bowman2020c}. The heat-engine mechanism is efficient in exciting low-radial order pressure and gravity modes with periods of order several hours \citep{Dziembowski1993f, Pamyat1999b}. Space photometry from the TESS mission has demonstrated that $\beta$~Cep stars are fairly common and there may be no upper mass limit for these stars \citep{Burssens2020a}. An example light curve of the $\beta$~Cep star HD~192575 \citep{Burssens2023a} is shown in Fig.~\ref{figure: light curves}. Historically, the high-amplitude pulsations observed using ground-based telescopes for a handful of $\beta$~Cep stars have proven extremely fruitful in constraining their interior rotation profiles (e.g. \citealt{Aerts2003d, Dupret2004b}). With modern space telescopes and more sophisticated forward asteroseismic modelling, the precision on mass, age, rotation and mixing that is achievable has much improved in recent years. For example, non-rigid rotation and a precise core mass and age were achieved for the $\beta$~Cep star HD~192575 thanks to modern TESS mission light curves \citep{Burssens2023a}.


    \subsection{SLF variability}
    \label{subsection: SLF}
    
    The now-retired {\it Kepler} and ongoing TESS space missions have revealed that almost all massive stars exhibit a non-periodic form of variability known as stochastic low-frequency (SLF) variability \citep{Bowman2019b}. It is characterised by a power excess in a frequency spectrum with amplitudes ranging from micro-magnitudes to milli-magnitudes, and a broad period range from several days to minutes. Although, the physical mechanism for SLF variability is not firmly established, there are a few non-mutually exclusive mechanisms: stochastically excited gravity waves (i.e. damped gravity modes with finite lifetimes) from the convective core; turbulence arising from partial ionisation zones (e.g. iron bump) for main-sequence massive stars with pre-dominantly radiative envelopes; and optically-thick and clumpy line-driven winds \citep{Bowman2023b}. Regardless of excitation mechanism, SLF variability has been shown to probe a massive star's mass and age \citep{Bowman2020b}, and has opened the window of performing wave asteroseismology of massive stars that do not have the high-amplitude heat-driven coherent pulsation modes.

    
    \subsection{Subdwarf stars}
    \label{subsection: sdB}

    The subdwarf B (sdB) stars are low-luminosity B-type stars with masses less than about 0.5~M$_{\odot}$, and are in an unusual stage of evolution because they have experienced significant mass loss as a red giant star to leave behind a helium core \citep{ASTERO_BOOK}. They are located between the red giant branch and the extreme horizontal branch in the HR~diagram and are progenitors of white dwarf stars. The discovery of their pulsational properties is fairly recent \citep{Kilkenny1997a}, with a growing number of variable sdB (i.e. sdBV) stars grouped into whether they are dominated by pressure or gravity modes. The gravity-mode sdB stars are cooler than the pressure-mode sdB stars, making them analogous to SPB and $\beta$~Cep stars, respectively. The pulsations in sdB stars are excited by the heat-engine mechanism, and are commonly low-radial order and low-angular degree \citep{Kilkenny2007b}. Asteroseismology of sdB stars has revealed tight constraints on their interior rotation \citep{Charpinet2018a}, as well as their masses and ages (e.g. \citealt{Charpinet2006c, Charpinet2019b, Zong2016b}).


    \subsection{White dwarf stars}
    \label{subsection: WDs}

    Eventually all low- and intermediate-mass stars become white dwarf stars, with the white dwarf cooling track shown in the HR~diagram in Fig.~\ref{figure: HRD}. Spectroscopic studies divide white dwarf stars into three main spectral classes of DO, DB, and DA going from hottest to coolest. The pulsating members of these spectral groups are referred to as GW~Vir stars, V777~Her stars, and ZZ~Cet stars, respectively \citep{ASTERO_BOOK}. Pulsations in white dwarf stars are excited by the heat-engine mechanism, although it is sometimes referred to as convective driving for DA white dwarf stars (e.g. \citealt{Brickhill1991a}). Pulsations in white dwarf stars are low-angular degree pressure and/or gravity modes based on their spectral type, and typically have short pulsation periods owing to their high densities --- see reviews by \citet{Fontaine2008e}, \citet{Winget2008} and \citet{Corsico2019a}. Asteroseismology of white dwarf stars has revealed their interior rotation \citep{Hermes2017f} and chemical composition profiles to great precision (e.g. \citealt{Giammichele2018a, Charpinet2019b}).


\section{Conclusions}
\label{section: conclusions}

Asteroseismology is a relatively new sub-field within stellar astrophysics, but it is growing rapidly in depth and breadth year on year. This advancement has been driven by the space photometry revolution in the last two decades, such as from the {\it Kepler} \citep{Borucki2010} and TESS \citep{Ricker2015} space telescopes, which have provided high-photometric precision and long-duration light curves for hundreds of thousands of stars. With improved data quality and quantity, which is orders of magnitudes better than the best cases previously achieved from the ground, asteroseismology has been applied to tens of thousands of red giant stars \citep{Chaplin2013c}, several hundred main-sequence pulsators and dozens of compact objects like white dwarf stars and subdwarf B~stars \citep{Kurtz2022a}. In so doing, asteroseismology has provided some of the most precise measurements of fundamental stellar parameters, such as masses, radii and ages, as well as accurate constraints on interior physical processes such as rotation and mixing mechanisms \citep{Aerts2021a}.

One of the most important conclusions of asteroseismology having been applied across the HR~diagram, including high-mass and intermediate-mass dwarf stars, evolved low-mass and intermediate-mass stars, as well as white dwarf stars and subdwarf B~stars, is that all stars have only a small amount of radial differential rotation \citep{Aerts2019b}. Moreover, the vast majority of stars for which asteroseismology using gravity-mode period-spacing patterns and/or rotational multiplets has revealed similar (near-)core and envelope rotation rates. Observed core-to-surface rotation ratios range approximately between 1-10 in thousands of red giant stars, whereas models predict at least an order of magnitude larger ratios than this for evolved stars \citep{Aerts2019b, Aerts2021a}. This means that a very efficient angular momentum transport mechanism, which remains challenging to explain \citep{Townsend2018a, Ouazzani2019a}, must be operating during most of a star's lifetime for a wide range of masses. This is because the full sample of asteroseismic interior rotation rates includes thousands of stars spanning the core-hydrogen, shell-hydrogen, core-helium evolutionary stages and beyond and all seem to show similar results \citep{Aerts2021a}. On the other hand, asteroseismic interior rotation profiles for massive stars are relatively few in number compared to low- and intermediate-mass stars, primarily because their rapid rotation, high binary fraction, and intrinsic scarcity in the Universe make them challenging to study \citep{Bowman2020c, Burssens2023a}. The strong missing angular momentum transport mechanism revealed by asteroseismology is an impactful conclusion for the wider astrophysics community. Inaccurate interior rotation profiles and unconstrained efficiencies in angular momentum transport mechanisms lead to uncalibrated interior chemical mixing processes in stellar evolution models, which limits our ability to reliably estimate stellar masses and ages. Leading explanations for the missing angular momentum transport mechanism(s) are magnetic fields and non-radial pulsations, both of which have shown great promise but are not mutually exclusive \citep{Aerts2019b}. 

In addition to time-series space photometry surveys, other impactful projects within astrophysics include ESA's GAIA mission \citep{Gaia2016} and large-scale ground-based spectroscopic campaigns, which are providing highly complementary data for constraining fundamental parameters. Ultra-precise parallaxes from the GAIA mission combined with accurate spectroscopic constraints on effective temperatures allows stars to be robustly placed in the HR~diagram, which greatly aids asteroseismic analysis. In the near future, ESA's PLATO mission, with a planned launch date in 2026 \citep{Rauer2024}, will greatly build on the previous {\it Kepler} and ongoing TESS missions by providing high-precision and long-duration light curves of millions of stars across the sky. The advantageous synergy between asteroseismology and the study of exoplanets, and the golden age of asteroseismology is expected to continue and reach new heights in the coming decades.

\begin{BoxTypeA}[box2]{Summary of Key Points}
    \begin{enumerate}
        \item Stellar pulsations are described in terms of spherical harmonic geometry, with a given pulsation mode frequency having a unique radial order ($n$), angular degree ($\ell$), and azimuthal order ($m$).
        \item Long-duration, high-precision, continuous time-series data are required to unambiguously identify pulsation mode frequencies, and measure diagnostics such as gravity-mode period spacing patterns, and rotationally- and magnetically-split pulsation multiplets.
        \item Pulsations are commonplace across the HR~diagram, and a diverse range of variability, including binarity, rotation, and pulsations allows asteroseismology to study the physics of stellar interiors.
        \item Asteroseismology provides precise constraints on interior physics of stars, such as rotation, mixing and magnetic fields, as well as fundamental parameters such as masses, radii and ages.
        \item Most stars are observed to have only a small amount of radial differential rotation, meaning an efficient angular momentum transport mechanism operates throughout stellar evolution that remains challenging to calibrate in stellar evolution models.
        \item Direct insight of the strength and geometry of magnetic fields deep inside main-sequence and evolved stars is now possible thanks to magneto-asteroseismology.
    \end{enumerate}
\end{BoxTypeA}


\begin{ack}[Acknowledgments]

DMB gratefully acknowledges funding from UK Research and Innovation (UKRI) in the form of a Frontier Research Grant under the UK government’s ERC Horizon Europe funding guarantee (SYMPHONY; grant number: EP/Y031059/1), and a Royal Society University Research Fellowship (URF; grant number: URF{\textbackslash}R1{\textbackslash}231631). The authors are grateful to C. S. Jeffrey for generously making Fig.~\ref{figure: HRD}. and members of the TASC community (\url{https://tasoc.dk}) for comments that improved the content of this review.
The {\it Kepler} and TESS data presented in this paper were obtained from the Mikulski Archive for Space Telescopes (MAST; \url{https://archive.stsci.edu/}) at the Space Telescope Science Institute (STScI), which is operated by the Association of Universities for Research in Astronomy, Inc., under NASA contract NAS5-26555. Support to MAST for these data is provided by the NASA Office of Space Science via the grants NAG5-7584 and NNX09AF08G, and other grants and contracts. Funding for the TESS mission is provided by the NASA Explorer Program, and funding for the {\it Kepler} and K2 missions was provided by NASA's Science Mission Directorate.

\end{ack}



\newcommand{\aap}{Astron. Astrophys.}
\newcommand{\mnras}{Mon. Not. Roy. Astron. Soc.}
\newcommand{\apj}{Astrophys. J.}
\newcommand{\apjs}{Astrophys. J. Sup. Series}
\newcommand{\apjl}{Astrophys. J. Letters}
\newcommand{\actaa}{Acta Astronomica}
\newcommand{\araa}{Ann. Rev. Astron. Astrophys.}
\newcommand{\aj}{Astron. Journal}
\newcommand{\apss}{Astrophys. Space Sci.}
\newcommand{\nat}{Nature}
\newcommand{\aapr}{Astron. Astrophys. Rev.}
\newcommand{\pasp}{Pub. Astron. Soc. Pacific}

\bibliographystyle{Harvard}
\bibliography{reference}

\end{document}